# A Beamforming Microwave Interferometric Radiometer for High-resolution Passive Imaging: Concept, Modeling, and Preliminary Demonstration

Ziyang Zhang, Hao Liu, Donghao Han, *Member, IEEE*, Te Wang, Jiyi Bian, Bingxu Li, Xing Tong, Mengyao Jiang

***Abstract*—High-resolution passive microwave imaging is important for numerical weather prediction, disaster monitoring, and oceanographic studies, but kilometer-level spatial resolution remains difficult to achieve because of aperture limitations and the high complexity of large interferometric arrays. This paper proposes a beamforming microwave interferometric radiometer (BF-MIR) for high-resolution passive microwave imaging. BF-MIR employs beamforming-capable antennas as interferometric elements in a large sparse array. The enlarged spatial-frequency sampling interval reduces the required number of elements and the cross-correlation burden, while a large aperture-to-sampling-interval ratio factor (ASRF) array design enables narrow-beam spatial filtering to suppress brightness temperature (TB) aliasing caused by spatial-frequency under sampling. In addition, beamforming enables dynamic beam steering across multiple pointing directions, thereby compensating for the limited instantaneous coverage of narrow beams. A beamforming interferometric imaging model is established, and the relationships among spatial resolution, radiometric sensitivity, and effective field of view are analyzed. An image-domain Shift-Accumulate method is further introduced to analyze aliasing, based on which an aliasing suppression strategy is developed. In addition, a three-element proof-of-concept prototype provides preliminary experimental validation of dynamic beam interferometric measurement and dynamic beam observation modes. These results indicate that BF-MIR is a promising architecture for further spaceborne high-resolution passive microwave imaging.**

## I. INTRODUCTION

Passive microwave remote sensing, which measures natural electromagnetic radiation from earth atmosphere and surface, plays a crucial role in numerical weather forecasting, climate prediction, environmental monitoring, and disaster management [1]. Conventional real-aperture radiometers achieve spatial resolution proportional to the antenna size; thus, high resolution requires large physical apertures with mechanical scanning, posing significant technical challenge for spaceborne platforms. Aperture synthesis techniques [2], pioneered by the Electronically Steered Thinned Array Radiometer (ESTAR) [3] and later implemented in the Microwave Imaging Radiometer using Aperture Synthesis (MIRAS) [4] aboard European Soil Moisture and Ocean Salinity (SMOS) Mission and Microwave Imager Combined Active and Passive (MICAP) [5] aboard Chinese Ocean Salinity Mission (COSM), overcome this limitation by using sparse arrays of small antennas to synthesize a large aperture. However, state-of-the-art Microwave Interferometric Radiometer (MIR) still offer resolutions on the order of tens of kilometers [2], considerably lower than those of existing optical and radar observations.

Studies on (sub)mesoscale oceanic processes, monitoring of mesoscale and small-scale weather systems, and observations of inhomogeneous land surfaces impose increasingly higher spatial resolution requirements on spaceborne microwave radiometry [6], [7]. In recent years, empowered by artificial intelligence, meteorological and oceanic numerical modeling systems have been rapidly advancing toward a global kilometer-scale resolution [8], [9], [10], [11]. Innovations in data assimilation methodologies have further enabled the complete and effective utilization of massive volumes of high-resolution observational data [12], [13]. In this context, improving the spatial resolution of spaceborne passive microwave observations to the kilometer scale level while maintaining adequate radiometric sensitivity, temporal coverage and feasibility, has thus become a technical challenge with clear practical significance.

Increasing spatial resolution requires a corresponding increase in aperture size for both real-aperture and synthetic-aperture systems. For real-aperture radiometers, higher resolution demands a larger antenna and faster scanning to maintain coverage, which reduces integration time and degrades radiometric sensitivity. Although multibeam reflector antennas may help relieve this tradeoff, they also increase system complexity [14]. The CIMR mission, designed for low earth orbit(LEO) operation, adopts an 8-m deployable reflector to achieve about 5-km resolution at 18.7GHz [15]. This

Ziyang Zhang, Jiyi Bian and Bingxu Li are with the Key Laboratory of Microwave Remote Sensing, National Space Science Center, Chinese Academy of Sciences, Beijing 100190, China, and also with the School of Electronic, Electrical and Communication Engineer, University of Chinese Academy of Sciences, Beijing 100049, China (e-mail: ziyang_zhang_0403@outlook.com).

Hao Liu, Donghao Han, Te Wang, Xing Tong and Mengyao Jiang are with the Key Laboratory of Microwave Remote Sensing, National Space Science Center, Chinese Academy of Sciences, Beijing 100190, China (Corresponding author: Hao Liu, e-mail: liuhao@mirslab.cn).

highlights a key challenge: further improving spatial resolution would impose substantially higher requirements on antenna size, scan mechanism design, and in-orbit reliability. Synthetic-aperture radiometers avoid a single ultra-large antenna but face rapid growth in array complexity and cross-correlation burden, together with sensitivity degradation caused by high array sparseness. Therefore, both technical routes encounter significant challenges when extended toward kilometer-level spaceborne passive microwave imaging.

Compared with Low Earth Orbit (LEO), Geostationary Earth Orbit (GEO) offers rapid revisit observation capabilities matched to high spatial resolution, which is particularly critical for the study of meso- and small-scale meteorological and oceanic processes. However, the orbital altitude of GEO at 36,000 km—45 times than that typical800km of LEO—requires a 45-fold increase in antenna aperture for the same spatial resolution, posing significant technical challenges. Existing GEO-oriented microwave interferometric radiometer studies, including NASA's GeoSTAR [16], [17], [18] and NSSC's GIMS [19], [20], [21], adopted larger interferometric elements to improve radiometric sensitivity, at the cost of a reduced instantaneous field of view. However, these systems have remained largely in the tens-of-kilometers resolution regime. At 18.7 GHz, achieving 10km, 5km, and 1km spatial resolutions from GEO requires Y-shaped array arm lengths of approximately 38m, 76m, and 378m, respectively, implying a rapidly increasing array scale and correlation complexity.

More recently, distributed formation-flying and on-orbit assembly technologies have provided viable approaches for realizing ultra-large antenna apertures in space [22], [23], [24], [25], [26], [27]. However, MIR imaging requires dense and continuous baseline sampling. When scaled toward higher spatial resolution, both technologies encounter two interrelated consequences: the number of interferometric elements increases proportionally with the aperture size, radiometric sensitivity degrades proportionally under a fixed integration time. Therefore, structural feasibility alone does not resolve the fundamental architecture challenge of kilometer-level passive microwave imaging. In this context, previous GEO instruments such as GIMS and GeoSTAR adopted enlarged interferometric elements (increased ASRF) to improve sensitivity and reduce number of interferometric elements, at the cost of narrower instantaneous field of view compensated by mechanical scanning. However, at kilometer-scale resolution, the required synthesized aperture reaches tens to hundreds of meters, making mechanical scanning practically infeasible due to structural constraints and reliability concerns.

In addition to aperture synthesis, phased-array technology has also attracted increasing attention in passive microwave remote sensing in recent years [27]. The booming development of next-generation satellite communication systems has driven rapid advances in low-cost, highly integrated, large-scale phased-array technology [28], [29], [30], offering flexible and agile beam-steering capabilities. This motivates a re-examination of instrument architectures that combine dynamic electronic beam steering with large-ASRF interferometric systems.

Against this background, this paper proposes a beamforming microwave interferometric radiometer (BF-MIR), in which large-aperture, beam-steerable antennas are used as interferometric elements. By combining enlarged interferometric elements with dynamic beam steering, BF-MIR provides a new architectural route for high-resolution passive microwave imaging, particularly for large sparse arrays and kilometer-level observations from geostationary orbit.

The main contributions of this paper are summarized as follows:

1) A BF-MIR architecture is proposed for high-resolution passive microwave imaging, in which large-ASRF beam-steerable antennas are employed as the interferometric elements, and a corresponding beamforming interferometric imaging model is established.
2) A quantitative system-level analysis framework is developed to characterize the relationships among key design parameters and major performance metrics, including spatial resolution, radiometric sensitivity, and effective FOV, thereby providing guidance for BF-MIR system design.
3) To address the aliasing issue inherently introduced by the enlarged $d_u$ , a Shift-Accumulate method is proposed for aliasing analysis, and an aliasing suppression strategy based on large-ASRF arrays is developed.
4) A three-element proof-of-concept prototype is built to provide preliminary experimental validation of the proposed architecture, including dynamic beam interferometric phase measurement and dynamic beam observation under wide-area normal scanning and selected-beam staring modes.

The remainder of this paper is organized as follows. Section II introduces the BF-MIR concept and imaging model and analyzes system performance. Section III investigates aliasing and presents the proposed suppression strategy. Section IV provides a preliminary experimental demonstration using a three-element prototype. Section V discusses system-level implications, implementation tradeoffs, and calibration challenges of BF-MIR, and Section VI concludes the paper.

## II. Beamforming Microwave Interferometric Radiometer

BF-MIR combines beamforming with interferometric imaging to achieve high-resolution passive microwave imaging. This section will detail the model of beamforming interferometric imaging, system concept and observation model of BF-MIR, and the relationship between system performance and parameters.

### *A. Beamforming Interferometric Imaging Model*

In MIR, the scene TB is reconstructed in the spatial domain from samples of the visibility function (VF) acquired in the spatial-frequency domain through interferometric measurements [31]. For an ideal system, the relationship between the visibility function $V(u,v)$ and the scene brightness temperature $T(\xi,\eta)$ can be expressed as

$$V(\mathrm{u,v}) = \frac{\sqrt{D_i D_j}}{4\pi} \iint_{\xi^2+\eta^2\le 1} \frac{T(\xi,\eta)}{\sqrt{1-\xi^2-\eta^2}} \cdot F_{ni}(\xi,\eta) F_{nj}*(\xi,\eta) \cdot \tilde{r}\left(-\frac{u\xi+v\eta}{f_0}\right) \cdot e^{-j2\pi(u\xi+v\eta)} d\xi d\eta \quad (1)$$

where $(u_{ij}, v_{ij}) = \left(\frac{X_i - X_j}{\lambda_0}, \frac{Y_i - Y_j}{\lambda_0}\right)$ denotes the spatial-frequency sampling position corresponding to the baseline formed by two interferometric elements located at $(X_i, Y_i), (X_j, Y_j)$. $T(\xi,\eta)$ is the scene TB distribution, $(\xi,\eta)$ are the direction-cosine coordinates of the source with respect to the antenna array plane. $F_n(\xi,\eta)$ is the normalized voltage pattern of the interferometric element. $\tilde{r}\left(-\frac{u\xi+v\eta}{f_0}\right)$ represents the spatial decorrelation function accounting for fringe-washing effects.

For MIRs, $d_u$ determines the alias-free field of view (AF-FOV). Once the $d_u$ exceeds the Nyquist limit, periodic replicas begin to fold into the unit circle, thereby causing aliasing in the reconstructed TB and reducing the AF-FOV. Fig. 1 illustrates this evolution for a Y-shaped array, where the solid circle denotes the unit-circle field of view, the dashed circles represent the periodic replicas, and the blue area is AF-FOV.

In BF-MIR, in order to realize a very large synthesized aperture with a manageable number of interferometric elements, $d_u$ must be enlarged to several tens or even hundreds of wavelengths. Under such a sampling regime, aliasing is no longer a marginal effect but an inherent feature of the imaging process.

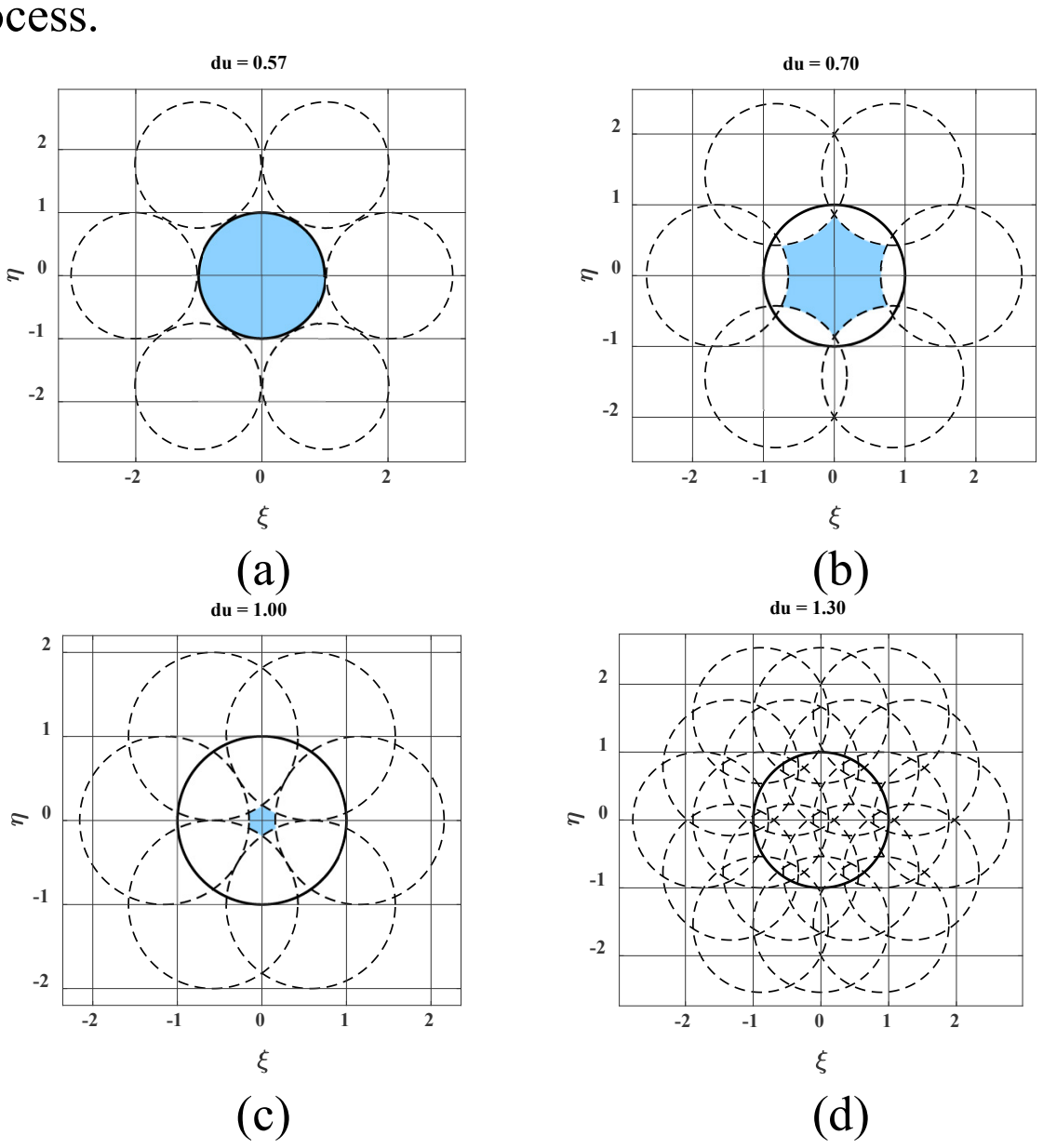

**Fig. 1.** Schematic diagram of aliasing for Y-shaped array. The solid circle represents the unit circle field of view, the dashed lines denote the replicas, and the blue area is AF-FOV. (a) $d_u$=0.57λ; (b) $d_u$=0.70λ; (c) $d_u$=1.00λ; (d) $d_u$=1.30λ.

To further illustrate the aliasing behavior of systems with large $d_u$, Fig. 2 presents a one-dimensional array example. Let $T_{scene}$ denote the scene TB. In Fig. 2 $T_{aliasing}$, the solid line represents the TB distribution within the unit circle, whereas the dashed lines correspond to the periodic replicas that fold into the unit circle, as illustrated in Fig. 1. The reconstructed TB within the unit circle, including the aliasing contributions, is denoted by $T_{rec}$.

As shown in Fig. 2(a), if the spacing between interferometric elements is directly enlarged while small-aperture wide-beam antennas are still used, the reconstructed TB is severely contaminated by aliasing. In contrast, as shown in Fig. 2(b), when the interferometric element aperture is increased with the $d_u$ such that a sufficiently large ASRF is maintained, the narrower antenna beam suppresses the replicas folding into the unit circle, thereby mitigating aliasing in the reconstructed TB.

This observation indicates that, in a BF-MIR system, the interferometric element serves not only as an energy transducer that converts free-space electromagnetic waves into guided signals in the receiver chain, but also as a spatial prefilter that exploits its antenna pattern to suppress aliasing in the reconstructed TB.

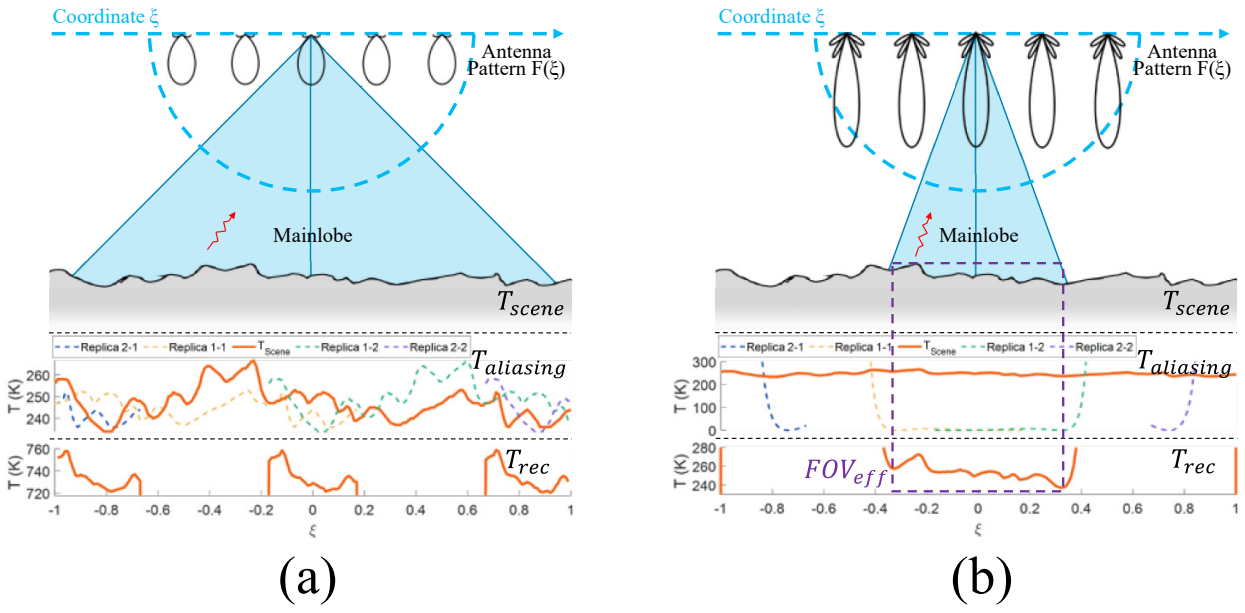

**Fig. 2.** Schematic diagram of aliasing in MIR with large $d_u$: (a) Interferometric elements with wide-beam antennas; (b) Interferometric elements with narrow-beam antennas.

As also illustrated in Fig. 2, when the $d_u$ is enlarged and the element main beam becomes narrower, the observation performance within the main-beam region can be maintained, but the instantaneous effective field of view (E-FOV) is reduced. To enlarge the imaging coverage, beam scanning can be introduced, either mechanically, as shown in Fig. 3(a), or electronically, as shown in Fig. 3(b).

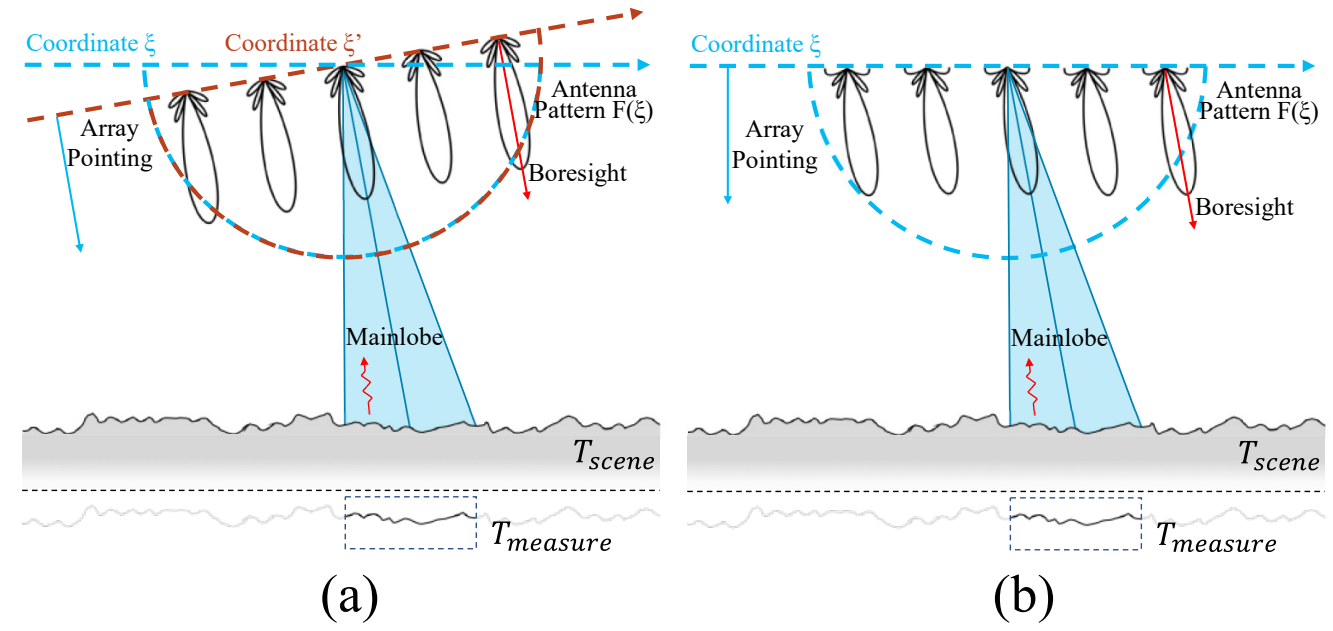

**Fig. 3**. Beam-scanning MIR imaging model. (a) Mechanical scanning; (b) Electronic scanning.

In this work, BF-MIR employs electronic beam scanning, as depicted in Fig. 3(b), to realize dynamic beam interferometric imaging. Both mechanical and electronic scanning can be used to expand the imaging FOV. Under mechanical scanning, the beam pointing direction remains aligned with the array normal. However, for a very large array, mechanical scanning is difficult to implement and generally provides a limited beam-switching rate. Under electronic scanning, the beam direction deviates from the array normal. Nevertheless, electromagnetic waves from the forward half-space can still impinge on the apertures of the interferometric elements. Therefore, the relationship between the reconstructed TB and the VF remains identical to that of a conventional MIR and is described by (2).

$$\tilde{T}(\xi,\eta) = \iint V(u,v)\cdot e^{j2\pi(u\xi+v\eta)}dudv \cdot \frac{4\pi\sqrt{1-\xi^2-\eta^2}}{\sqrt{D_iD_j}F_{ni}(\xi,\eta)F_{nj^*}(\xi,\eta)} \quad (2)$$

Since incident waves from sidelobe directions are attenuated by the antenna pattern of the interferometric elements, only the TB within the main-beam region can be effectively reconstructed. When the beam is steered toward $(\xi_0,\eta_0)$, the E-FOV TB of BF-MIR can be expressed as

$$\tilde{T}_E(\xi,\eta;\xi_0,\eta_0) = \tilde{T}(\xi,\eta); \quad (\xi-\xi_0)^2+(\eta-\eta_0)^2 \le FOV_E^2 \quad (3)$$

where $FOV_E \le 1-\sqrt{\xi_0^2+\eta_0^2}$ is the radius of the E-FOV.

*B. Instrument Concept*

Based on the beamforming interferometric imaging model established above, three representative architectural implementations of BF-MIR system are prosed, as illustrated in Fig. 4. These architectures use different types of beamforming interferometric elements: phased-array antennas(PAA), digital beamforming arrays (DBF), and phased-array-fed reflector antennas (PAF).

In a PAA-based BF-MIR, each interferometric element is realized as a phased-array antenna. The beamforming network combines signals from individual element antennas with phase shifts to form a directional beam. In a DBF-based architecture, each element antenna is digitized, and beamforming is performed digitally after analog-to-digital conversion. In a PAF architecture, the beamforming function is realized by the quasi-optical arrangement of a reflector and its feed array, where the reflector acts as the primary focusing element and the feed array provides phase-controlled illumination.

These architectures differ in structural complexity, beamforming capability, and electrical system complexity. PAA and DBF are planar structures that can be directly integrated on a spacecraft panel or deployed as flat arrays, offering a compact and low-profile form factor. In contrast, PAF requires a reflector antenna, which introduces additional mechanical complexity in deployment, alignment, and thermal management. In terms of beamforming capability, PAA can generate only a single beam at a time, requiring time-multiplexed switching to cover multiple beam positions. DBF and PAF, on the other hand, can support simultaneous multibeam operation at the hardware level, enabling parallel observation of multiple directions. Regarding electrical system complexity, PAA technology is relatively mature with well-established beamforming networks, offering moderate complexity. DBF is the most complex, as it requires digitization at the element or subarray level and high-throughput digital processing. PAF is the simplest: the beamforming function is largely realized by the quasi-optical feed arrangement and reflector geometry, requiring only a modest number of phase-controlled feed elements, leading to the lowest power consumption.

Despite these differences in antenna form and beamforming implementation, all three architectures share the same beamforming interferometric imaging principle: beamforming is first performed at the interferometric-element level, after which the beamformed outputs are correlated to obtain the visibility functions for image formation. The block diagrams in Fig. 4 emphasize this common signal flow.

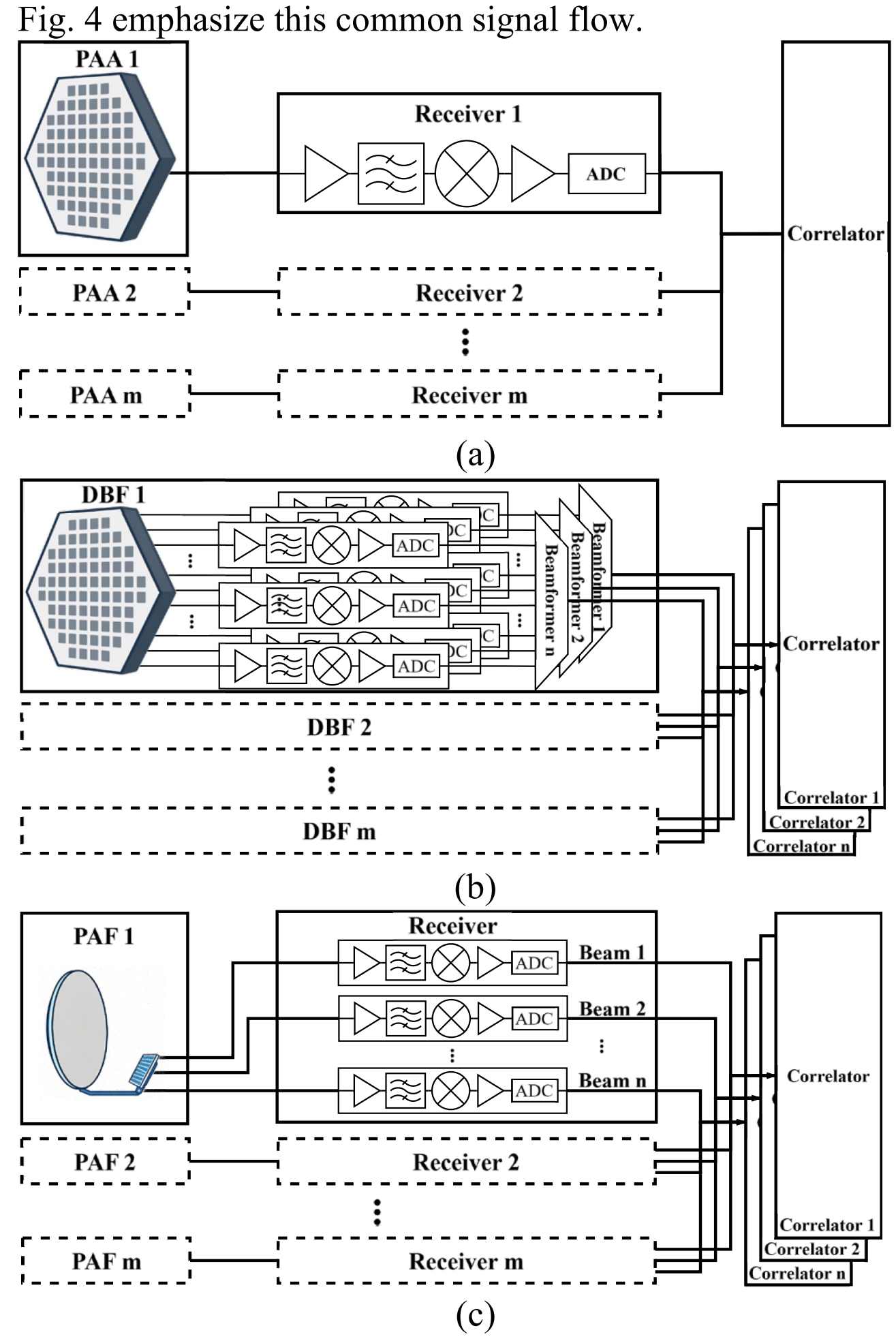

**Fig. 4.** Representative architectures of the BF-MIR system with three types of beamforming interferometric elements. (a) Phased-array antennas; (b) Digital beamforming arrays; (c) Phased-array-fed reflector antennas.

Based on this common principle, the proposed BF-MIR architecture introduces three interconnected features that together address the key design tradeoffs of high-resolution passive microwave imaging. The proposed BF-MIR architecture is applicable to different orbital configurations and observation scenarios. For clarity of presentation, the following discussion is developed using GEO Earth observation as a representative example.

**(1) Enlarged spatial-frequency sampling interval:**

By using large-aperture beamforming antennas as interferometric elements, BF-MIR can realize a larger synthesized aperture with fewer elements, thereby reducing hardware complexity and correlation burden while increasing the effective collecting area. The associated tradeoff is that the enlarged $d_u$ violates the Nyquist criterion, causing replicas to fold into the fundamental domain, as indicated by the red hexagon in Fig. 5, and produce aliasing in the reconstructed TB. Meanwhile, the enlarged interferometric-element aperture narrows the element beam, as illustrated by the half-power beam region (HPBR) in Fig. 5. Since the subsequent quantitative assessments are mainly meaningful within the main-beam-covered region, this paper defines the HPBR as the region used for sensitivity and aliasing-error evaluation.

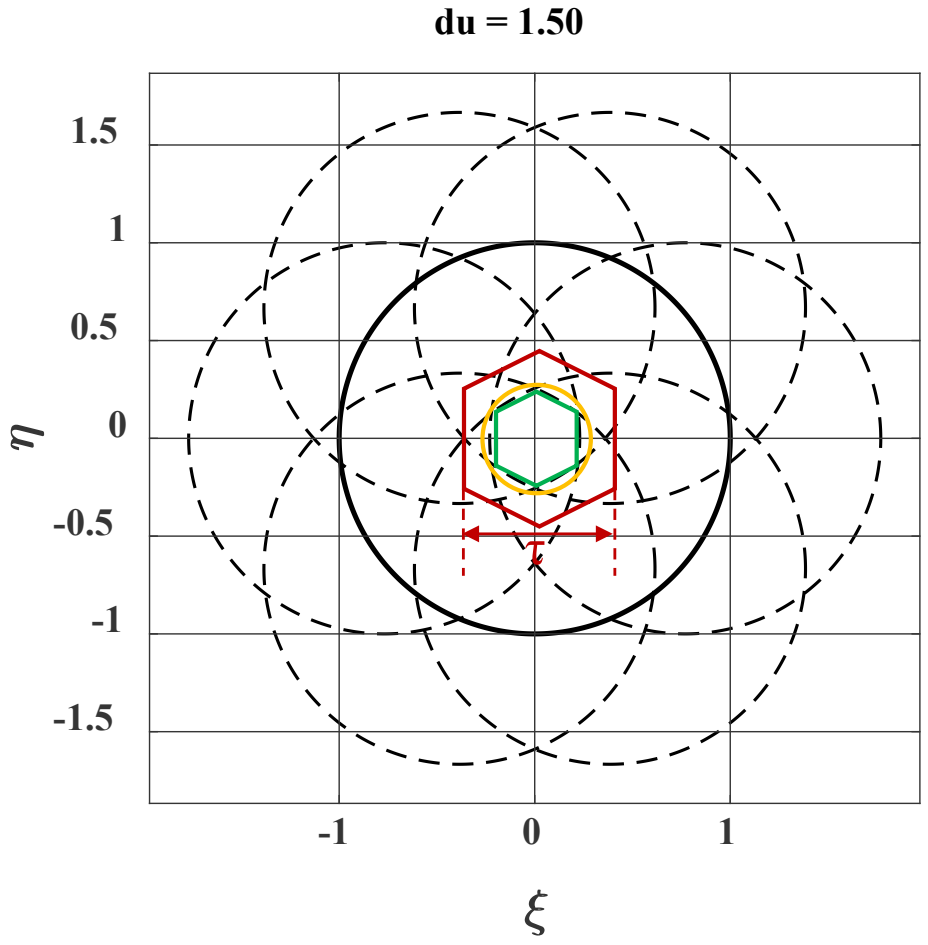

**Fig. 5.** The red hexagon is fundamental domain, the orange circle is half-power beam region (HPBR) of interferometric element, the green hexagon is the effective field of view.

**(2) Large-aperture-to-sampling-interval ratio factor array design:**

To suppress the aliasing introduced by the enlarged $d_u$, a multi-row array with a large ASRF is adopted. Fig. 6 illustrates the array configurations and corresponding spatial-frequency domain sampling points for GEO Earth observation with 5-km spatial resolution under different ASRF values. As ASRF increases from 1 to 4, the interferometric element transitions from a single-row to a multi-row layout, and the element aperture grows proportionally with $d_u$. This narrows the element beam and suppresses the aliasing contributions from replicas, while also increasing the total effective aperture area for improved radiometric sensitivity. However, the narrower beam reduces the instantaneous imaging FOV. Fig. 5 illustrates the relationship among the fundamental domain (red hexagon, determined by $d_u$), the 3-dB beam area (orange circle, determined by ASRF), and the effective field of view (green hexagon, the region where both conditions are satisfied).

**(3) Dynamic beam steering:**

The beamforming capability of each interferometric element enables agile beam steering to compensate for the coverage reduction caused by the narrow element beam. BF-MIR can support multiple observation modes, as shown in Fig. 7. In the normal observation mode of Fig. 7(a), the beams are steered over a predefined observation grid to cover a large area. In the simultaneous staring-and-tracking mode of Fig. 7(b), one beam continuously stares at a selected region, while another beam is steered for tracking observation. In addition, beamforming provides further flexibility in controlling the element radiation pattern, which can also be exploited for aliasing suppression.

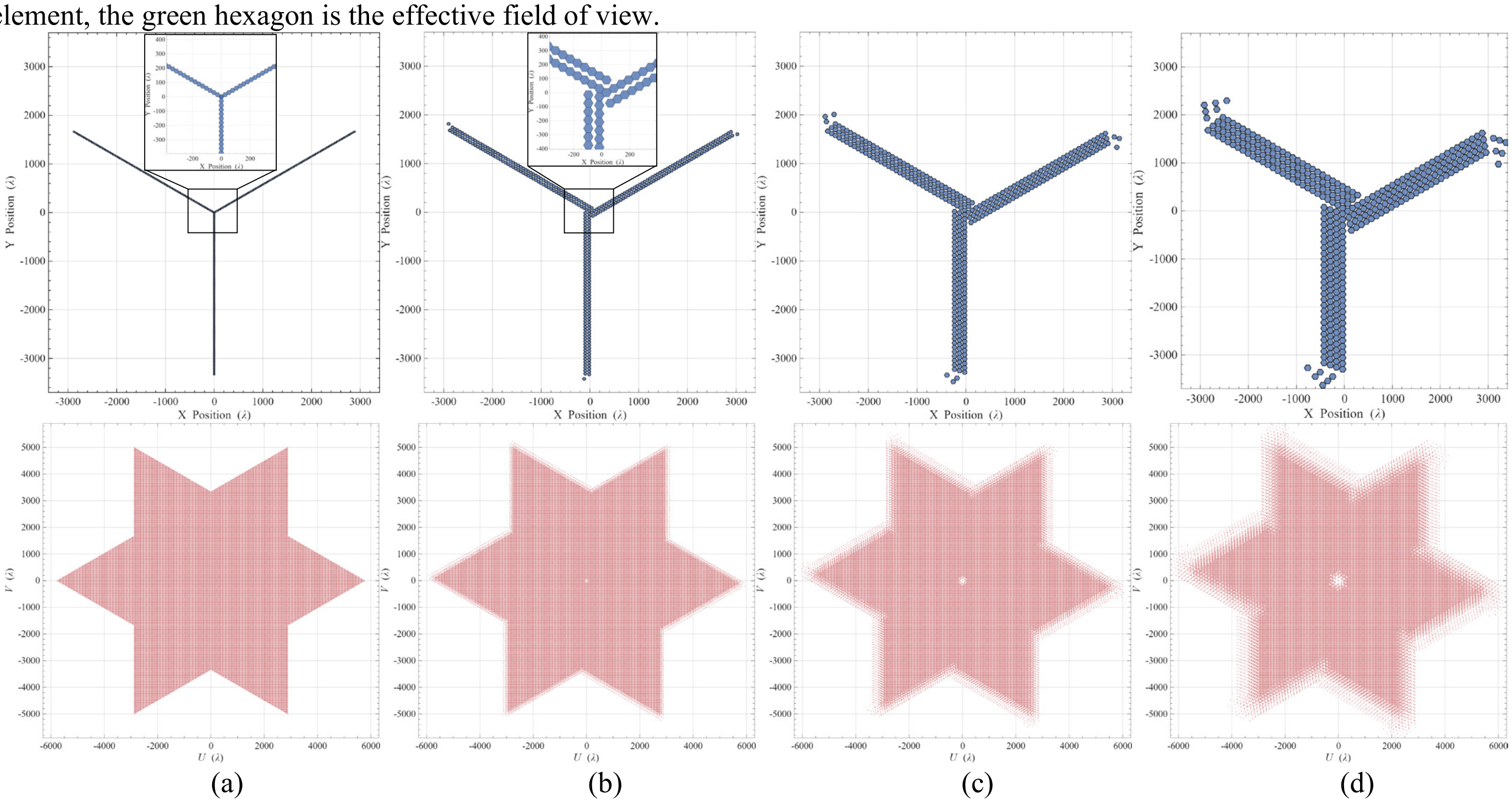

**Fig. 6.** Interferometric array configurations (top row) and spatial frequency domain sampling points (bottom row) for GEO Earth observation with 5-km spatial resolution when $d_u = 30$. (a) ASRF = 1; (b) ASRF = 2; (c) ASRF = 3; (d) ASRF = 4.

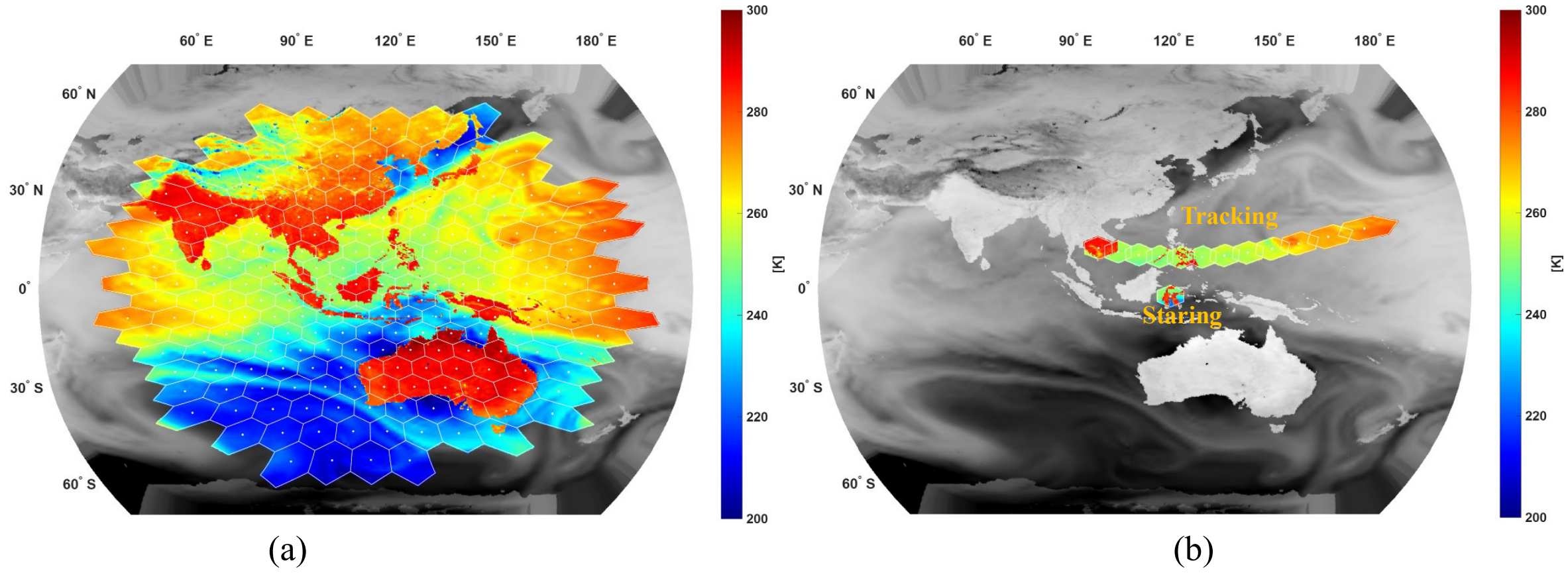

(a) (b)

**Fig. 7.** Schematic of BF-MIR observation modes: (a) Normal observation mode; (b) Simultaneous staring-and-tracking mode. The white hexagonal denote the E-FOV.

*C. System performance*

Based on the above analysis of the BF-MIR architecture and its characteristics, this subsection further establishes a performance-evaluation framework for BF-MIR and quantitatively relates its key system parameters to the main observation capabilities. Such relationships provide guidance for determining the critical design parameters according to application requirements and for conducting system-level engineering assessment. In the following, the spatial resolution, radiometric sensitivity, and effective FOV are discussed in detail in the context of the large $d_u$, large-ASRF array design, and dynamic beam steering adopted by BF-MIR.

**(1) Spatial Resolution**

The spatial resolution of MIR characterizes the minimum spatial detail that can be resolved in the reconstructed TB and is determined primarily by the overall size of the interferometric array. It is commonly quantified by the 3-dB width of the point spread function (PSF). As indicated by (2), the imaging relationship of BF-MIR is identical to that of a conventional MIR. Therefore, its PSF can be expressed as[1]

$$AF(\xi-\xi',\eta-\eta') = \Delta s \sum W(u,v)\cdot e^{j2\pi[u(\xi-\xi')+v(\eta-\eta')]} \tilde{r}\left(-\frac{u\xi+v\eta}{f_0}\right) \quad (4)$$

If the fringe-washing effect is neglected, the spatial resolution of the Y-shaped array with $N_i$ interferometric elements spaced $d_u$ are [32]

$$\Delta\xi = 1.41\cdot\frac{\alpha_{sw}}{d_u(N_i-1)} \quad (5)$$

where $\alpha_{sw}$ is the windowing factor. In the following system-level calculations, Hann windowing is applied to the visibility samples. The spatial-resolution broadening introduced by this visibility-domain tapering is accounted for by using a windowing factor of $\alpha_{sw}=1.4$.

A higher spatial resolution requires a larger interferometric aperture. To more directly illustrate the relationship between system complexity and spatial resolution, Fig. 8 shows, for a geostationary Earth-observation scenario, the number of interferometric elements required as a function of $d_u$ for target resolutions of 1 km, 5 km, and 10 km.

For high-resolution BF-MIR with large synthesized aperture, the fringe-washing effect can no longer be neglected and must be taken into account in system design. The fringe-washing function describes the spatial decorrelation caused by differences in the receiver frequency responses of two channels and by signal delay. Assuming identical receiver frequency responses, the fringe-washing effect can be written as

$$\tilde{r}\left(-\frac{u\xi+v\eta}{f_0}\right) = \operatorname{sinc}\left(-\frac{u\xi+v\eta}{f_0}\cdot B\right) \quad (6)$$

where $f_0$ represents the center frequency of the receiver chain, and B is the bandwidth.

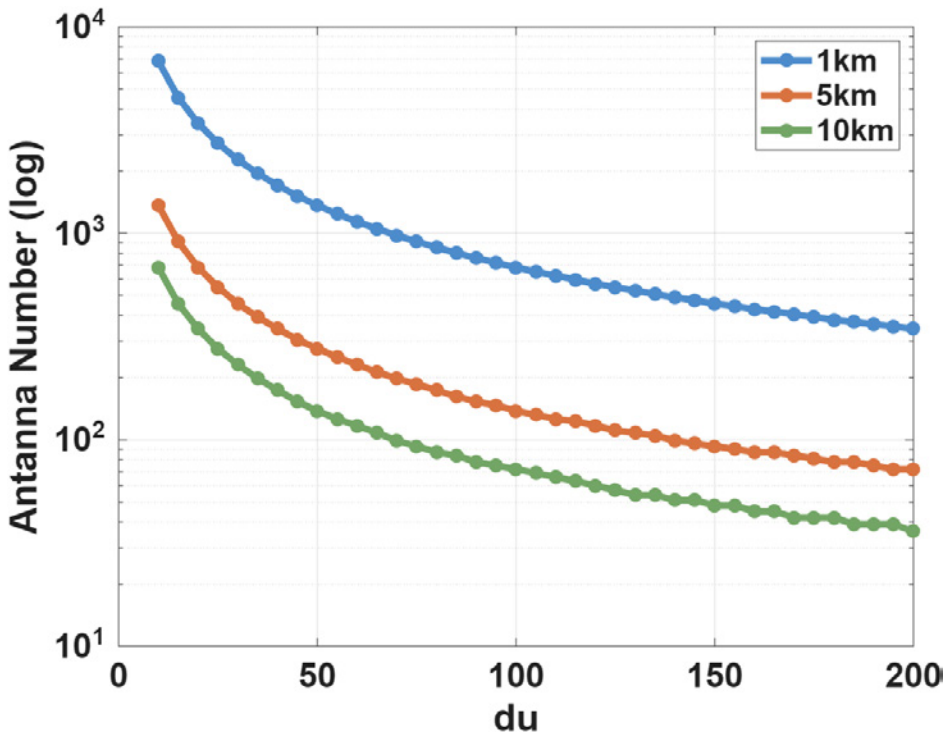

**Fig. 8.** Relationship between interferometric elements number and $d_u$ under different spatial resolutions for GEO Earth observation.

In general, a larger bandwidth, a larger array aperture, or a larger observation angle results in stronger fringe washing. To mitigate the fringe-washing effect in BF-MIR systems with large synthesized apertures, sub-band processing is employed[33], [34]. This approach divides the system bandwidth into multiple narrower sub-bands and performs correlation processing separately within each sub-band, thereby reducing the signal bandwidth seen by each correlator and suppressing spatial decorrelation, and the required sub-band bandwidth can be determined by

$$\left|\operatorname{sinc}\left(B_{sub}\frac{D\sin\theta}{c}\right)\right| \geq \rho \quad (7)$$

where $B_{sub}$ represents the sub-band bandwidth, c is the speed of light, D denotes the interferometric array aperture, and θ denotes the observation angle. The parameter $\rho$ denotes a prescribed decorrelation tolerance, representing the minimum acceptable normalized fringe-washing amplitude within the evaluation region. A larger $\rho$ corresponds to a stricter requirement and therefore leads to narrower sub-bands or a larger number of sub-bands.

For a BF-MIR configuration targeting 5-km geostationary Earth observation, setting the allowable decorrelation loss to 0.2 dB requires the sub-band bandwidth to be below 2.68 MHz at 18.7 GHz and 7.21MHz at 50.3 GHz, so that fringe-washing-induced decorrelation remains within the prescribed tolerance. Although sub-band processing increases the backend computational burden, the overall processing complexity remains within practical limits because the number of interferometric elements is substantially reduced in BF-MIR and because of the continuous advances in high-speed digital signal processing and on-board processing technologies [35], [36], [37].

Therefore, in the design of a high-resolution BF-MIR system, the required spatial resolution can be used to evaluate the corresponding system complexity, sub-band bandwidth, and related design constraints based on the above analysis.

**(2) Radiometric Sensitivity**

The radiometric sensitivity of an MIR characterizes the minimum TB variation that can be detected by the observation system. The sensitivity of the reconstructed modified TB can be expressed as [32]

$$\Delta T' = \frac{T_A+T_R}{\sqrt{B\tau}} \cdot \Delta s \cdot \sqrt{N_V} \cdot \alpha_W \tag{8}$$

where the antenna temperature and the equivalent receiver noise temperature are represented by $T_A$ and $T_R$ respectively, B is the system bandwidth, $\tau$ is the integration time, $\Delta$s is the area of the spatial-frequency sampling grid, $N_v$ is the number of nonredundant sampling points in the spatial-frequency domain, and $\alpha_w$ denotes the windowing factor that accounts for the influence of the reconstruction window on sensitivity.

Assuming that all interferometric elements have identical antenna patterns, the sensitivity of the reconstructed TB can be further obtained from (2) as

$$\Delta T_B = \frac{4\pi}{D} \cdot \frac{\sqrt{1-\xi^2-\eta^2}}{|F(\xi,\eta)|^2} \cdot \frac{T_A+T_R}{\sqrt{B\tau}} \cdot \Delta s \cdot \sqrt{N_V} \cdot \alpha_W \tag{9}$$

Unlike a conventional MIR, BF-MIR employs beam steering, and therefore the influence of beam pointing on radiometric sensitivity must be taken into account. Assuming that each interferometric element is realized as a planar phased-array antenna, the directional factor of the interferometric element for a beam steered toward $(\xi_0, \eta_0)$ can be written as

$$\begin{aligned} D(\xi_0,\eta_0) &= D(0,0)\sqrt{1-\xi_0^2-\eta_0^2} \\ &= \eta_\alpha \pi^2 (APF \cdot du)^2 \cdot \sqrt{1-\xi_0^2-\eta_0^2} \end{aligned} \tag{10}$$

where $\eta_\alpha$ is the antenna efficiency at array normal, $APF \cdot d_u$ is the diameter of the interferometric element in a large-ASRF array. For a fixed $d_u$, the directional factor increases with the ASRF and decreases as the beam is steered away from the array normal. Its maximum value is achieved when the beam points toward the array boresight.

The power pattern of an interferometric element under beam steering can be expressed as

$$|F(\xi,\eta;\xi_0,\eta_0)|^2 = |g(\xi,\eta)|^2 \cdot |AF(\xi-\xi_0,\eta-\eta_0)|^2 \tag{11}$$

where $|g(\xi,\eta)|^2$ is the element power pattern and $AF(\xi,\eta)$ denotes the array factor. For a phased-array antenna, the element pattern is usually assumed to be omnidirectional and therefore independent of beam steering, whereas the array factor shifts with the beam-pointing direction without broadening and without changing its intrinsic shape.

Substituting (10) and (11) into (9), the radiometric sensitivity of the reconstructed TB for BF-MIR can be obtained as

$$\begin{aligned} \Delta T_B(\xi,\eta;\xi_0,\eta_0) = & \frac{1}{APF^2} \frac{1}{\sqrt{1-\xi_0^2-\eta_0^2}} \frac{\sqrt{1-\xi^2-\eta^2}}{|AF(\xi-\xi_0,\eta-\eta_0)|^2} \\ & \frac{4\pi}{\eta_\alpha \pi^2 du^2} \cdot \frac{T_A+T_R}{\sqrt{B\tau}} \cdot \Delta s \cdot \sqrt{N_V} \cdot \alpha_W \end{aligned} \tag{12}$$

Equation (12) shows that, similar to a conventional MIR, the BF-MIR achieves its highest sensitivity near the beam center, and its sensitivity can also be improved by increasing the bandwidth or the integration time. In addition, BF-MIR can further improve sensitivity by increasing the ASRF. However, its radiometric sensitivity degrades as the beam pointing moves away from the array normal.

Based on the above analysis, Fig. 9 shows, for the geostationary Earth-observation case, the sensitivity at the array boresight as a function of spatial resolution, $d_u$, and ASRF. The integration time is set to 100ms.

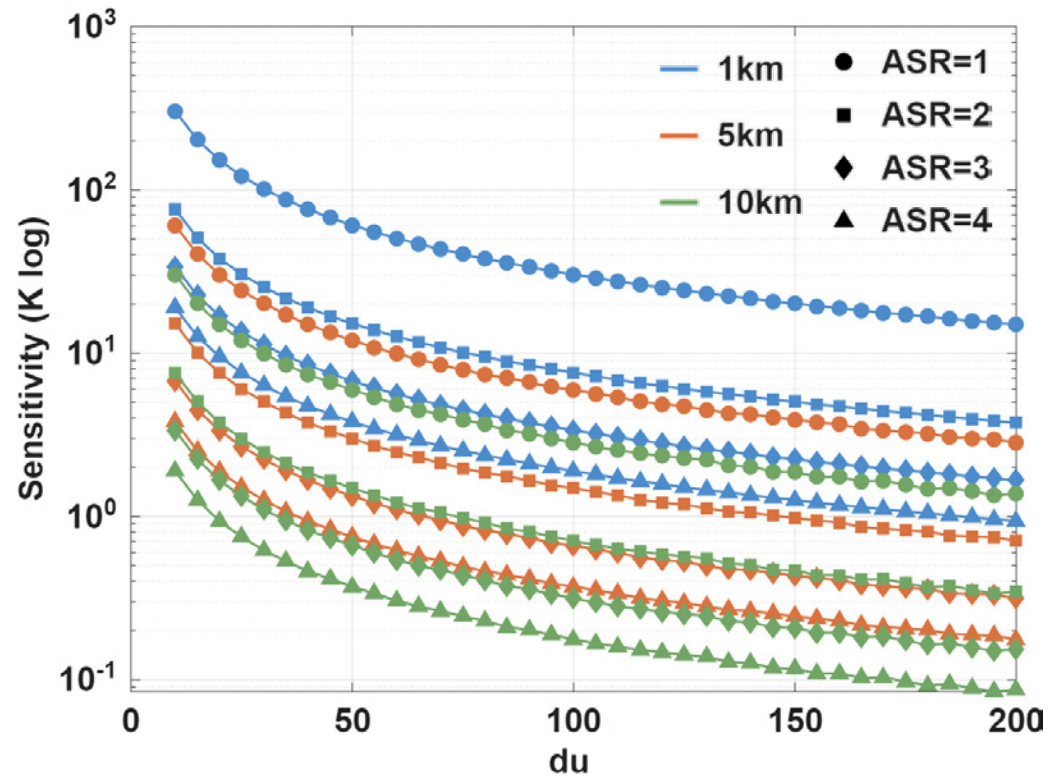

**Fig. 9.** Relationship between sensitivity and $d_u$ under different spatial resolutions and ASRFs for GEO Earth observation.

**(3) Effective Field of View**

In an MIR, the scene TB is reconstructed in the spatial domain from samples of the visibility function acquired in the spatial-frequency domain. The $d_u$ in the spatial-frequency domain determines the AF-FOV of the reconstructed TB. Since BF-MIR adopts a much larger $d_u$, replicas inevitably fold into the fundamental domain over the entire unit-circle FOV. Consequently, an absolutely AF-FOV no longer exists for BF-MIR.

To characterize the imaging coverage of BF-MIR under this condition, the E-FOV is introduced. Because aliasing cannot be completely eliminated, the imaging FOV should instead be defined according to the residual aliasing level achievable through system design. As shown in Fig. 5, the E-FOV is defined as the region that satisfies both of the following conditions: 1) it is located within the 3-dB beamwidth of the interferometric element, and 2) its root-mean-square (RMS) aliasing error is below a prescribed threshold. This threshold should be selected according to the observation target and application requirements. A detailed analysis of the E-FOV, based on the aliasing characteristics of BF-MIR, is present in Section III.

Overall, the parameter design of BF-MIR reflects a fundamental tradeoff between high spatial resolution and engineering feasibility. As shown by Fig. 7 and Table Ⅰ, enlarging $d_u$ can substantially reduce the number of interferometric elements required for a given target resolution, whereas increasing ASRF improves collecting area at the cost of a larger and more complex interferometric element. For GEO Earth observation, the system scale grows rapidly as the target resolution is pushed from 10 km to 1 km, indicating that kilometer-level passive microwave imaging requires a careful balance among array aperture, element count, element complexity, sensitivity, and effective FOV. At the same time, the large-$d_u$ regime that makes such scaling feasible also introduces severe aliasing in the reconstructed TB, which directly limits the E-FOV. Therefore, the next section focuses on the aliasing mechanism in large-spacing systems and on the corresponding suppression strategy.

Table Ⅰ

SYSTEM-SCALE PARAMETERS FOR GEO EARTH OBSERVATION

| Resolution | Arm length | du | $N_i$ | $N_e$ under different ASRF | | |
|---|---|---|---|---|---|---|
| | | | | 1 | 2 | 3 |
| 1km | 378m@18.7GHz 140m@50.3GHz | 50λ | 1413 | 230 | 918 | 2066 |
| | | 100λ | 711 | 918 | 3673 | 8264 |
| | | 150λ | 474 | 2066 | 8264 | 18595 |
| 5km | 76m@18.7GHz 28m@50.3GHz | 30λ | 474 | 83 | 331 | 744 |
| | | 60λ | 240 | 331 | 1322 | 2975 |
| | | 90λ | 159 | 744 | 2975 | 6694 |
| 10km | 38m@18.7GHz 14m@50.3GHz | 20λ | 356 | 37 | 147 | 331 |
| | | 40λ | 180 | 147 | 588 | 1322 |
| | | 60λ | 123 | 331 | 1322 | 2975 |

Note: $N_i$ is the number of interferometric elements; $N_e$ is the number of element antenna per interferometric element with spacing of 3.3λ.

## III. ALIASING UNDER LARGE SAMPLING INTERVALS

BF-MIR reduces system complexity by enlarging the spatial-frequency sampling interval. However, once $d_u$ exceeds the Nyquist limit, aliasing becomes unavoidable in the reconstructed TB. Fig. 1 illustrates this effect for a Y-shaped array, where increasing $d_u$ causes periodic replicas in the spatial domain to fold into the fundamental domain. According to the Fourier-transform relationship between the visibility function and the reconstructed modified TB, the reconstructed TB in the presence of aliasing can be expressed as

$$\hat{T}_B(\xi,\eta) = \sum_{n=1}^{N} T_B^{'}(\xi-\xi_n,\eta-\eta_n)\frac{4\pi}{\sqrt{D_iD_j}} \cdot \frac{\sqrt{1-\xi^2-\eta^2}}{F_{ni}(\xi,\eta)\cdot F_{nj}*(\xi,\eta)} \quad (13)$$

where $T_B'$ denotes the modified TB including aliasing, and $(\xi_n,\eta_n)$ represents the location of the nth replica. For arrays whose spatial-frequency samples lie on a hexagonal grid, the replica locations can be written as

$$(\xi_n,\eta_n) = \left(\left(l+\frac{k}{2}\right)\cdot\frac{2}{\sqrt{3}du},\frac{k}{du}\right) \quad k,l = 0,\pm1,\pm2\ldots \quad (14)$$

For systems with large $d_u$, the effective imaging FOV cannot exceed the fundamental domain determined by $d_u$. As $d_u$ increases, the fundamental domain becomes smaller. According to (13) and Fig. 1(d), the aliasing behavior of large-spacing systems is governed mainly by two factors: 1) the number of replicas that fold into the fundamental domain, and 2) the aliasing contribution of each replica.

The number of replicas contributing to aliasing is determined by $d_u$, because $d_u$ defines the periodicity of the replicas and therefore determines how many replicas overlap the fundamental domain. In contrast, the aliasing contribution of an individual replica is jointly determined by its relative position with respect to the unit circle and by the antenna-pattern weighting. The relative position determines which portion of the replica folds into the fundamental domain, whereas the antenna pattern determines how strongly that portion is weighted. Together, these two factors determine the aliasing strength of the replica entering the fundamental domain.

Based on the above considerations, this section is organized as follows. In Part A, a Shift-Accumulate method is proposed for analyzing aliasing in large-spacing systems. In Part B, this method is used to investigate the effect of the $d_u$ on aliasing. In Part C, the influence of antenna-pattern characteristics, including the ASRF and weighting scheme, on aliasing is analyzed. Finally, in Part D, an aliasing suppression strategy is developed based on the above analysis and validated through simulations.

### A. Shift-Accumulate Method

Conventional simulation methods for MIRs are typically based on inverse Fourier transformation or G-matrix inversion and are generally suitable for evaluating the overall imaging behavior of MIR systems [38], [39]. However, when applied to systems with large $d_u$, these methods tend to mix the aliasing caused by periodic replicas with other effects, such as the windowing effect introduced by finite spatial-frequency sampling. As a result, it is difficult to isolate and quantitatively evaluate the contribution of a specific periodic replica to the imaging error.

To analyze the aliasing characteristics of BF-MIR, this paper proposes a Shift-Accumulate method based on (13). The proposed method directly simulates the aliasing process in the image domain, thereby enabling analytical separation of the aliasing contribution from other system nonidealities. The implementation procedure is summarized as follows.

1) Modified TB generation: Based on the high-resolution original TB scene$T_B(\zeta,\eta)$ shown in Fig. 10(a) which corresponds to the region covered by the staring observation mode in Fig. 7(b), and the power pattern of the interferometric element $|F_n(\zeta,\eta)|^2$, the modified TB $T_B'(\zeta,\eta)$ is first calculated, as shown in Fig. 10(b).
2) Replica localization: According to $d_u$, the center position of the nth periodic replica $(\zeta_n,\eta_n)$ to be evaluated is determined in the direction-cosine domain.

For a hexagonal sampling grid, the replica location is given by (14).

3) Aliasing accumulation: The portion of the selected periodic replica that folds into the fundamental domain is extracted and then added to the corresponding position in the fundamental domain. In this way, the modified TB $\hat{T}'_B(\zeta,\eta)$ including the accumulated aliasing contributions is obtained. As an example, Fig. 10(c) shows the modified TB after accumulation of the six replicas in the first ring.
4) Reconstructed TB calculation: The aliased modified TB is divided by the antenna pattern over the corresponding unit-circle region to obtain the reconstructed TB containing aliasing $\hat{T}_B(\zeta,\eta)$, as illustrated in Fig. 10(d).

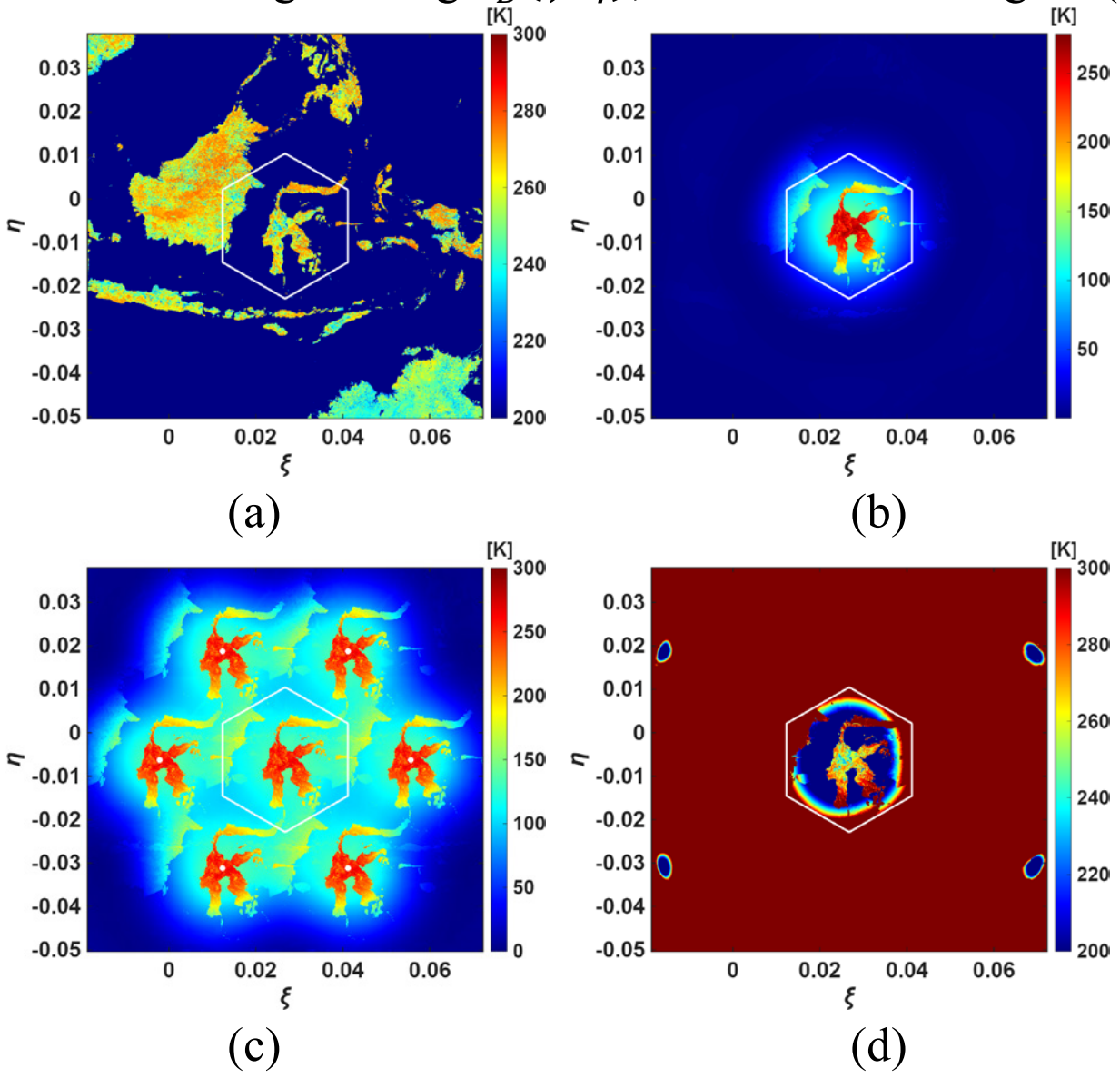

**Fig. 10.** Shift-Accumulate method. (a) Input original TB; (b) Modified TB; (c) Modified TB with aliasing; (d) Reconstructed TB with aliasing.

This method is particularly suitable for analyzing the aliasing behavior of individual periodic replicas and for revealing the dominant factors governing aliasing in large-spacing BF-MIR systems. Since the Shift-Accumulate analysis is performed in the image domain, it is intended to isolate aliasing contributions caused by spatial-frequency under sampling rather than to reproduce the complete visibility-sampling and image-reconstruction process of a finite array. Consequently, the maximum baseline length and the corresponding nominal spatial resolution are not the primary controlling parameters in this analysis; instead, the aliasing behavior is mainly governed by the spatial-frequency sampling interval $d_u$, the antenna pattern determined by ASRF and weighting, and the selected evaluation region.

Since the Shift-Accumulate method performs aliasing simulation in the image domain, the evaluation of kilometer-level high-resolution imaging systems requires original TB data with hectometer-level spatial resolution. However, no global TB dataset with such resolution is currently available. Therefore, pseudo-TB data with similar spatial characteristics are adopted in this work. For land scenes, 250-m-resolution pseudo-TB data are generated from the global normalized difference vegetation index (NDVI). For ocean scenes, because the TB varies more smoothly, high-resolution ocean TB data are obtained by interpolating low-resolution ocean TB data. Finally, the land and ocean pseudo-TB data are combined to form a global pseudo-TB dataset with hectometer-level resolution.

### *B. Effect of Sampling Interval on Aliasing*

Under a fixed ASRF, enlarging $d_u$ reduces the number of interferometric elements and the associated cross-correlation burden. However, it also decreases the spacing between periodic replicas in the spatial domain, potentially allowing more replicas to overlap the fundamental domain. This subsection examines whether increasing $d_u$ leads to a substantial worsening of aliasing and, consequently, whether $d_u$ should be regarded as a primary aliasing-control parameter in BF-MIR design. To this end, the Shift-Accumulate method is used to analyze aliasing from two perspectives: 1) the contributions of individual aliasing zones, and 2) the overall aliasing produced by all periodic replicas.

As shown in Fig. 11, the replicas are distributed radially on a hexagonal lattice in the direction-cosine domain. Since replicas located at similar radial distances from the center of the unit circle experience similar antenna-pattern attenuation, they tend to make comparable contributions to the aliasing error. They are therefore grouped into the same aliasing zone, defined as

$$(n-1)\cdot\tau < r \le n\cdot\tau \tag{15}$$

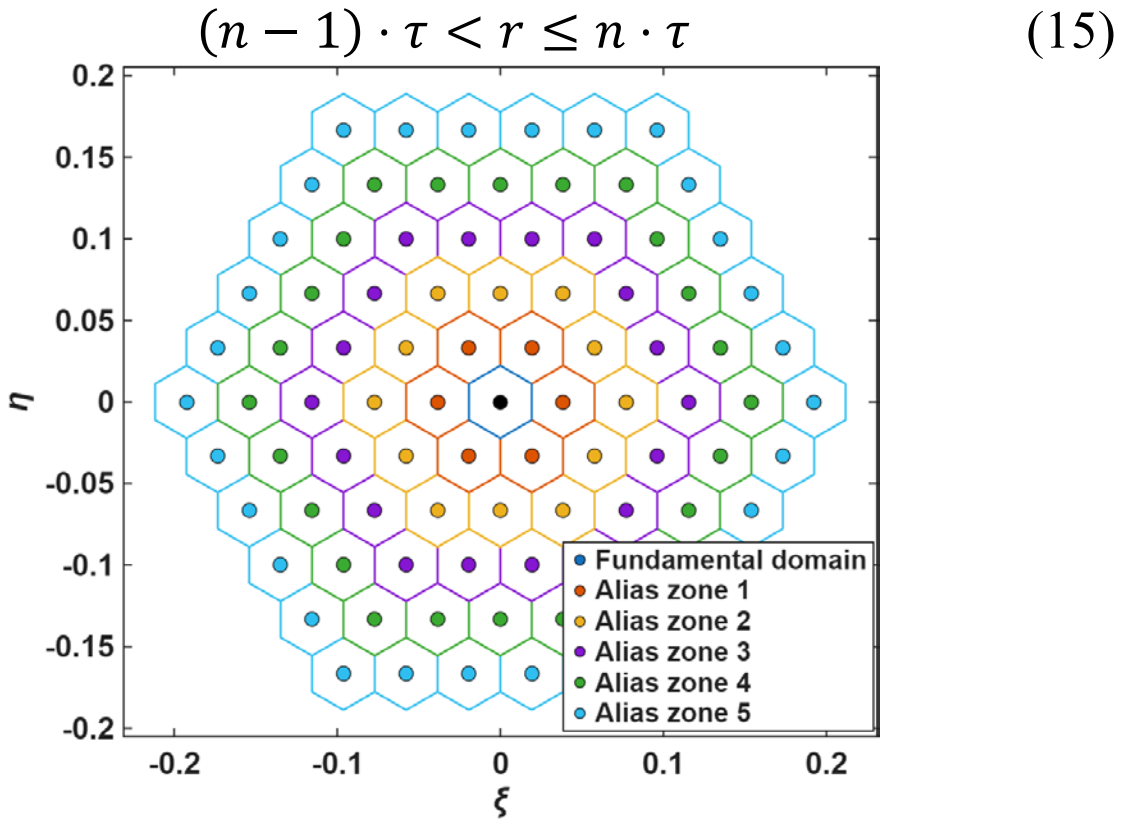

**Fig. 11.** Partitioning of aliasing zones when $d_u$=30λ.

To isolate the intrinsic effect of $d_u$ from scene-dependent spatial variability, the original scene is set to a uniform TB of 300 K in the Shift-Accumulate simulations. The antenna pattern of each interferometric element is modeled as that of a uniformly excited hexagonal phased-array antenna whose opposite-side distance is ASRF times $d_u$. The array elements are arranged on a triangular lattice with an element spacing of 3.3 λ, and the element pattern is assumed to be that of a uniformly excited circular aperture antenna with a diameter of 3.3 λ. The HPBR of the interferometric element, defined in Section II-B and illustrated in Fig. 5, is used as the evaluation region for calculating the RMS aliasing error. This choice ensures that the aliasing assessment is performed over the main-beam-covered area where BF-MIR provides effective observations.

The results in Fig. 12 (a)reveal that the aliasing contribution decreases rapidly as the aliasing-zone index increases. Although more distant zones contain a larger number of replicas, these replicas are farther from the antenna main-beam region and are therefore more strongly attenuated by the antenna pattern. As a result, their contributions to the aliasing error within the HPBR become progressively weaker. This indicates that the nearest few aliasing zones dominate the aliasing behavior and should therefore receive primary attention in BF-MIR design.

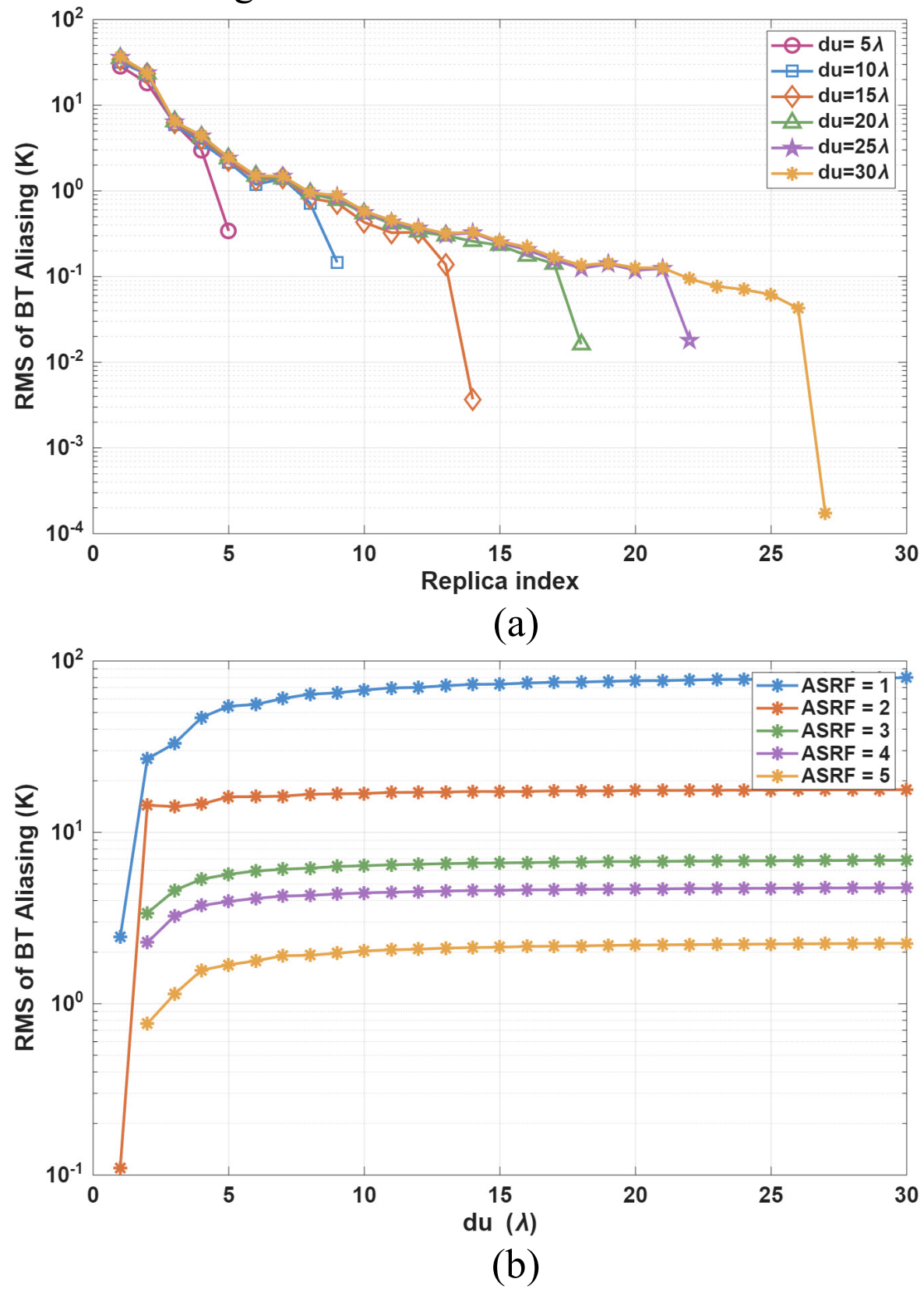

**Fig. 12.** (a) Aliasing of individual aliasing zones under different $d_u d_u$ when ASRF=1; (b) Aliasing of all aliasing zones under different $d_u$ and ASRFs.

Fig. 12(a) also shows that, under a fixed ASRF, the aliasing contribution of each individual zone varies only slightly with $d_u$. As $d_u$ increases, the replicas associated with the same aliasing zone move closer to the center of the unit circle, which tends to increase their geometric overlap with the fundamental domain. At the same time, because the ASRF is kept constant, the physical aperture of the interferometric element increases proportionally with $d_u$, leading to a narrower beam and stronger attenuation of the replicas. These two effects largely offset each other, making the contribution of an individual aliasing zone only weakly dependent on $d_u$.

This behavior explains the trend observed in Fig. 12(b), where the overall aliasing produced by all periodic replicas increases only slightly with $d_u$ and then gradually saturates. A larger $d_u$ allows more aliasing zones to overlap the fundamental domain, but the additional contribution from newly involved distant zones is increasingly small because their energy is strongly suppressed by the antenna pattern. Consequently, although the total aliasing level rises moderately at first, it does not exhibit substantial deterioration as $d_u$ continues to increase.

The above results indicate that, under a fixed ASRF, enlarging $d_u$ has only a limited impact on the aliasing level within the main-beam evaluation region. Therefore, $d_u$ may be selected primarily according to system-complexity considerations, including element count and correlation burden, without causing a significant worsening of aliasing. This also implies that further aliasing suppression in BF-MIR should rely mainly on antenna-pattern design rather than on the sampling interval itself, which is examined in the next subsection.

### *C. Effects of Antenna-Pattern*

Since Section III-B shows that the sampling interval $d_u$ has only a limited effect on aliasing under a fixed ASRF, further aliasing suppression in BF-MIR must rely mainly on the antenna-pattern characteristics of the interferometric elements. In this subsection, two pattern-related factors are examined: 1) beam narrowing associated with the ASRF, and 2) sidelobe control introduced by antenna weighting. Their influences are evaluated using both quantitative RMS aliasing errors and representative image-domain results.

Using the Shift-Accumulate method introduced in Section III-B, two sets of simulations are conducted to evaluate the influence of antenna-pattern characteristics on aliasing. The first set is a quantitative simulation based on a uniform 300 K scene, which is used to isolate the intrinsic effect of antenna-pattern design from scene-dependent spatial variations. The same array configuration, element model, and evaluation-region definition as those in Section III-B are adopted, while the ASRF and antenna weighting scheme are varied. The RMS aliasing errors calculated within the evaluation region are summarized in Table II. The second set is a representative image-domain simulation using a structured TB scene under a GEO fixed-beam observation geometry. In this case, the beam is directed toward the Kalimantan region (111.8°E, 0°), with $d_u = 30\lambda$, and the same array and element models are used. The corresponding reconstructed TB images and aliasing-error distributions are shown in Fig. 13. Therefore, Table II provides quantitative comparisons of aliasing levels under different ASRF and weighting configurations, whereas Fig. 13 illustrates the spatial characteristics of aliasing in a representative observation scene.

TABLE II

RMS ALIASING LEVEL IN THE EVALUATION REGION

| ASRF | Uniform (K) | Taylor -25dB (K) | Taylor -30dB (K) | Hamming (K) |
|---|---|---|---|---|
| 1 | 111.31 | 177.62 | 234.58 | 428.65 |
| 2 | 14.011 | 4.48 | 0.86 | 1.76 |
| 3 | 5.253 | 2.11 | 0.54 | 0.14 |
| 4 | 3.617 | 1.44 | 0.43 | 0.10 |

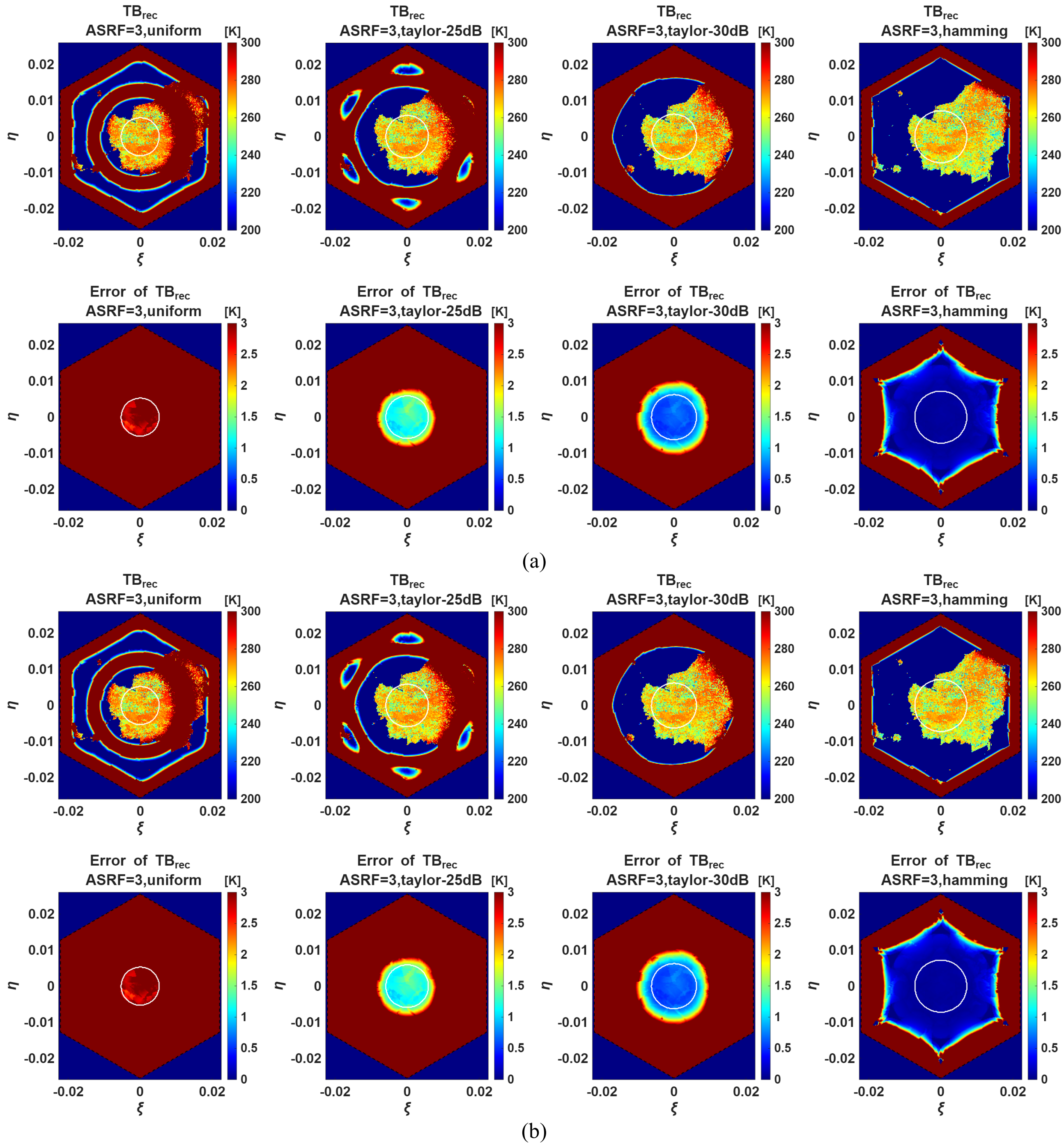

**Fig. 13.** Representative image-domain aliasing results for a GEO fixed-beam observation scene over the Kalimantan region (111.8°E,0°), $d_u$=30 λ . The reconstructed brightness temperature and corresponding aliasing error are shown under different ASRF and weighting configurations. The white circles indicate the HPBR of the interferometric element. (a) Taylor (-25dB) weighting with ASRF from 1 to 3; (b) Uniform weighting, -25dB Taylor weighting, -30dB Taylor weighting and hamming weighting for ASRF=3.

**(1) Effect of the ASRF**

As evidenced jointly by the RMS aliasing errors in Table II and the image-domain results in Fig. 13 (a), the aliasing level decreases markedly as the ASRF increases. For all weighting schemes listed in Table II, increasing the ASRF from 1 to 2 leads to a substantial reduction (e.g., from 111.31 K to 14.01 K for uniform weighting), while further increasing the ASRF continues to improve the aliasing performance but with diminishing returns. This trend occurs because a larger ASRF narrows the element beam, which suppresses the main beam contributions from the nearest replicas that dominate the aliasing error when the beam is wide. Once the nearest replica contributions are sufficiently attenuated, the residual aliasing is increasingly governed by weaker sidelobe contributions, so additional beam narrowing provides smaller incremental gains. The image domain results in Fig. 13(a) confirm this behavior: the reconstructed TB becomes progressively cleaner as ASRF increases, and the HPBR boundary shrinks accordingly.

**(2) Effect of the Antenna Weighting**

The influence of weighting depends strongly on whether sufficient beam narrowing has already been achieved. When the ASRF is small (ASRF = 1), weighting does not necessarily reduce aliasing effectively and may even increase the RMS aliasing level, because the residual aliasing is still dominated by the broad main-beam contribution of the nearest replicas. In contrast, when the ASRF becomes sufficiently large (≥ 2), the dominant residual aliasing is no longer controlled by those main-beam components but mainly by energy entering through the sidelobes. In this regime, lower-sidelobe weighting becomes genuinely effective. Table II shows that at ASRF = 4, for example, the RMS error drops from 3.617 K (uniform) to 0.10 K (Hamming weighting). The image-domain comparison in Fig. 13(b) further illustrates this: for ASRF = 3, reducing the sidelobe level yields a cleaner central imaging region. However, weighting also reduces aperture efficiency, which lowers the usable aperture gain. Therefore, antenna weighting should be regarded only as a secondary optimization tool that becomes beneficial after sufficient beam narrowing has been established, rather than as an independent primary suppression mechanism.

**(3) Residual Aliasing and Design Implications**

Another important observation from the image-domain results is that the residual aliasing is spatially nonuniform within the beam-covered region: the aliasing error is weakest near the FOV center and increases toward the beam edge. Consequently, the usable imaging region depends on both the adopted evaluation region and the application dependent aliasing tolerance. The effective field of view (E-FOV) is therefore interpreted as a qualitative, requirement-dependent concept rather than a fixed threshold-based metric.

Based on the above analysis, a practical design sequence for aliasing control in BF-MIR can be established. First, the spatial-frequency sampling interval $d_u$ is selected according to system-level complexity constraints — such as element count and cross-correlation burden — because its influence on aliasing is limited when the ASRF is fixed. Second, the ASRF is chosen to provide the primary aliasing suppression through sufficient beam narrowing. After the main-beam contribution of the nearest replicas has been adequately attenuated, moderate antenna weighting can be introduced as a secondary optimization to further suppress sidelobe-dominated residual aliasing. This design sequence reflects a tradeoff among aliasing suppression, instantaneous imaging coverage, and aperture efficiency: increasing ASRF reduces aliasing but decreases the beam-limited coverage for a single pointing direction; weighting reduces residual aliasing but lowers aperture efficiency. The E-FOV, being requirement-dependent, should be evaluated in the context of this tradeoff.

In addition to these system-design measures, further mitigation of aliasing may also be possible at the data-processing stage through model-based reconstruction, de-aliasing, regularized inversion, or other signal-processing methods[40], [41]. Such processing-oriented approaches may further relax the tradeoff between aliasing suppression and usable coverage and will be investigated in future work.

## IV. Preliminary Demonstration

This section presents a preliminary proof-of-concept demonstration of the proposed BF-MIR architecture using a three-element prototype. The experiments are intended to validate key operational functions required by BF-MIR, rather than to reproduce the performance of a full-scale high-resolution imaging system. Specifically, the first experiment is conducted in a microwave anechoic chamber using a point-source target to demonstrate the interferometric phase-measurement capability of phased-array-based interferometric elements under beam-steering operation. The second experiment is performed under real outdoor observation conditions to demonstrate dynamic beam observation modes corresponding to the normal scanning and selected-beam staring operations introduced in Section II-B. In addition, a radiometric sensitivity check is conducted to characterize whether the large-aperture beamforming elements provide sufficient measurement stability for the subsequent dynamic observation experiment.

The prototype consists of three identical planar phased-array antennas used as beamforming-capable interferometric elements, as shown in Fig. 14. These phased-array antennas are mature commercial off-the-shelf (COTS) products. Their main parameters, including array scale, operating frequency, bandwidth, beamwidth, and steering range, are lisfted in Table III. The beamformed outputs of the three antennas are received by independent channels and cross-correlated to obtain the visibility samples of the three baselines.

TABLE III

Prototype Parameter

| Parameter | Value |
|---|---|
| **Frequency** | 18.7GHz |
| **Bandwidth** | 200MHz |
| **Integration time** | 10ms |
| **3-dB beamwidth** | 3.4° @0° |
| **Elevation scan range** | 0~60° |
| **Azimuth range** | 0~360° |
| **Element antenna number** | 1024 |

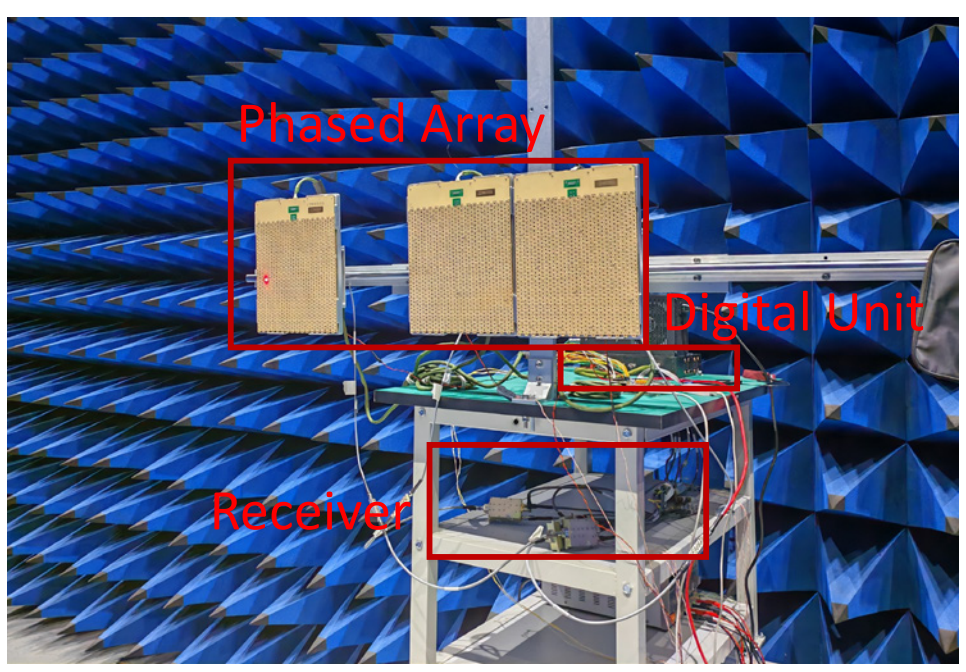

**Fig. 14.** Three elements BF-MIR prototype.

### A. Dynamic Beam Interferometric Experiment

Using the prototype described above, a dynamic beam interferometric phase measurement experiment was first conducted. In a conventional MIR, interferometric measurements are usually performed under a fixed antenna beam. For a point-like target moving at a constant velocity, the phase of the complex cross-correlation varies approximately linearly with the target position along the baseline direction. This linear phase behavior is a fundamental characteristic of the interferometric fringe and provides a direct way to verify the interferometric measurement capability of the prototype.

For BF-MIR, however, the interferometric elements are beamforming-capable antennas, and the beam direction may change during observation. Therefore, it is necessary to verify whether consistent cross-correlation phases can still be obtained under beam-steering operation. In particular, electronic beam steering introduces beam-state-dependent phase offsets among the interferometric elements. These offsets are caused by residual channel mismatch, phase-control errors, and nonideal consistency among the phased-array antennas, and must be compensated before the measured phase can be interpreted as the interferometric fringe phase.

When a point target is observed by BF-MIR, the phase of the cross-correlation measured by the baseline formed by the i-th and j-th interferometric elements can be expressed as

$$\phi(i,j;\xi_0,\eta_0;\xi,\eta) = \phi_1(i,j;\xi,\eta) + \phi_2(i,j;\xi_0,\eta_0) = 2\pi\left(u_{ij}\xi + v_{ij}\eta\right) + \phi_2(i,j;\xi_0+\eta_0) \quad (16)$$

where $\phi_1(i,j;\xi,\eta)$ denotes the cross-correlation phase of the i,j baseline for a point target located at $(\xi,\eta)$. The term $\phi_2(i,j;\xi_0,\eta_0)$ represents the phase offset between the i-th and j-th elements when the beam is steered toward $(\xi_0,\eta_0)$. This phase offset is caused by the nonideal consistency among the interferometric elements and varies with the beam-pointing direction. In this experiment, the phase offsets corresponding to different beam positions were measured in advance and compensated for before analyzing the fringe phase.

As illustrated in Fig. 15, a broadband noise source was mounted on a two-dimensional scanning platform and moved at a constant speed along the x-axis. During the measurement, the beam direction was electronically adjusted in steps of 1.5° so that the noise source remained within the central region of the antenna beam. At each beam position, the complex cross-correlations of the three baselines were recorded. After beam-state-dependent phase-offset compensation, the cross-correlation phase was compared with the expected linear fringe response.

The measured results are shown in Fig. 15. The compensated cross-correlation phase varies approximately linearly with the target position, which is consistent with the theoretical interferometric fringe behavior. The phase error shown in Fig. 15 remains within ±1.5°, indicating that the residual measurement error after compensation is small compared with the overall fringe-phase variation. These results demonstrate that the prototype can maintain consistent interferometric phase measurements during beam-steering operation.

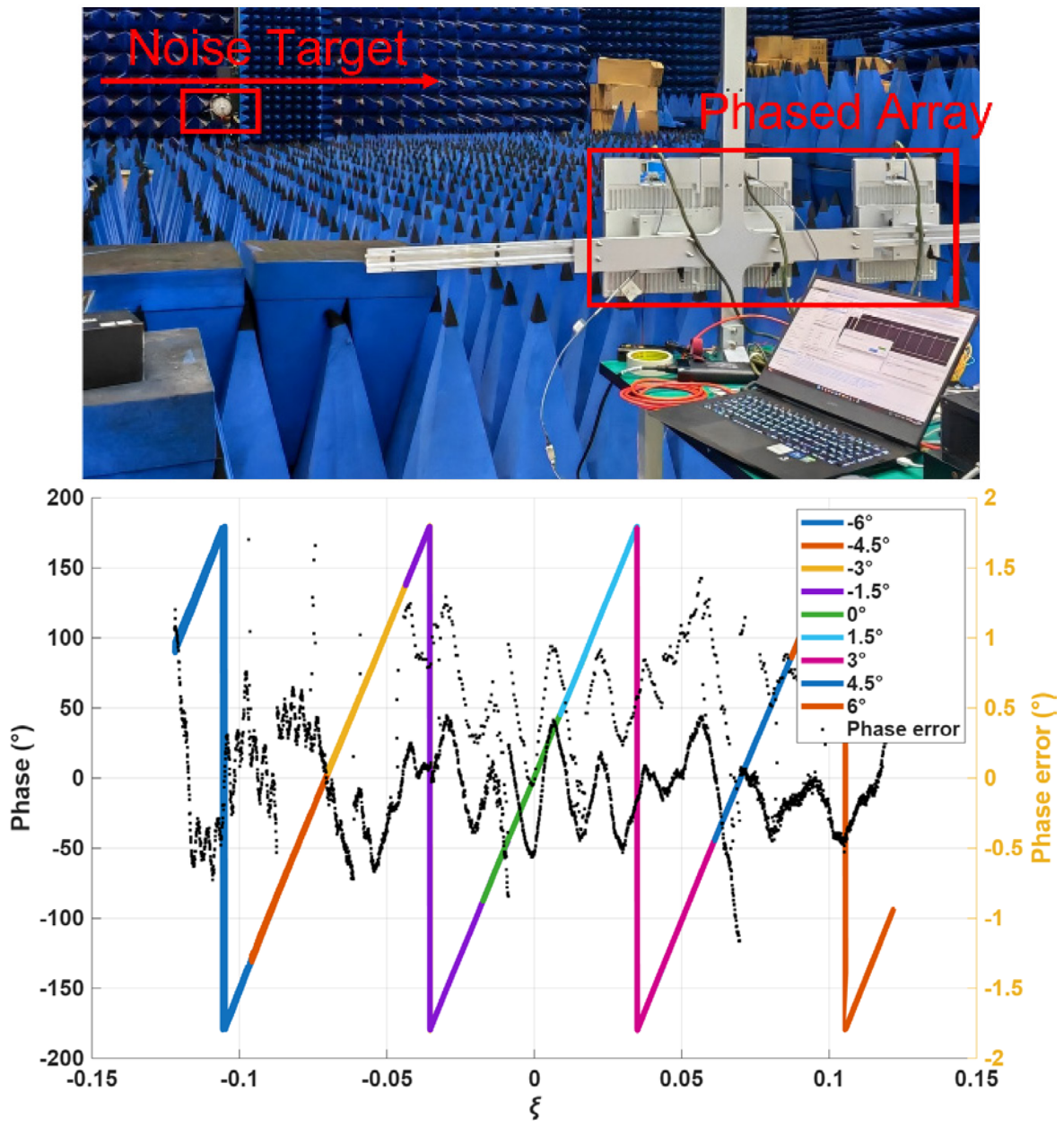

**Fig. 15.** Interference Fringe Testing.

This experiment provides preliminary validation of dynamic beam interferometric phase measurement in BF-MIR. Together with the dynamic beam observation experiment presented in the next subsection, it supports the feasibility of using beamforming-capable antennas as interferometric elements for dynamically steered passive microwave interferometric observation.

### B. Validation of Dynamic Beam Observation Modes

In addition to the dynamic interferometric phase measurement demonstrated above, BF-MIR also requires the capability to electronically switch among multiple beam directions and acquire interferometric measurements under different observation modes. This subsection presents a dynamic beam observation experiment using the three-element prototype under natural-field experimental conditions. The experiment includes two representative operation modes: wide-area normal scanning and repeated staring over selected beam positions, which correspond to the spaceborne observation modes introduced in Section II-B. Before these dynamic observations, a radiometric sensitivity check is performed as a supporting characterization of the prototype.

Before the dynamic observation experiment, a sensitivity check was performed as a supporting characterization of the prototype. The measured noise figure of the receiver chain was approximately 7dB, corresponding to an equivalent receiver noise temperature of about 1500K. With a system bandwidth of 200MHz and an integration time of 10ms, the theoretical single-channel radiometric sensitivity is approximate 1.06K. Hot/cold calibration measurements were then carried out by alternately observing liquid nitrogen and a blackbody target, as shown in Fig. 16. The measured single-channel sensitivity was approximately 1.1 K, which is consistent with the theoretical estimate. This result confirms that the prototype has sufficient short-term radiometric stability for observing temporal variations in the subsequent dynamic beam experiment.

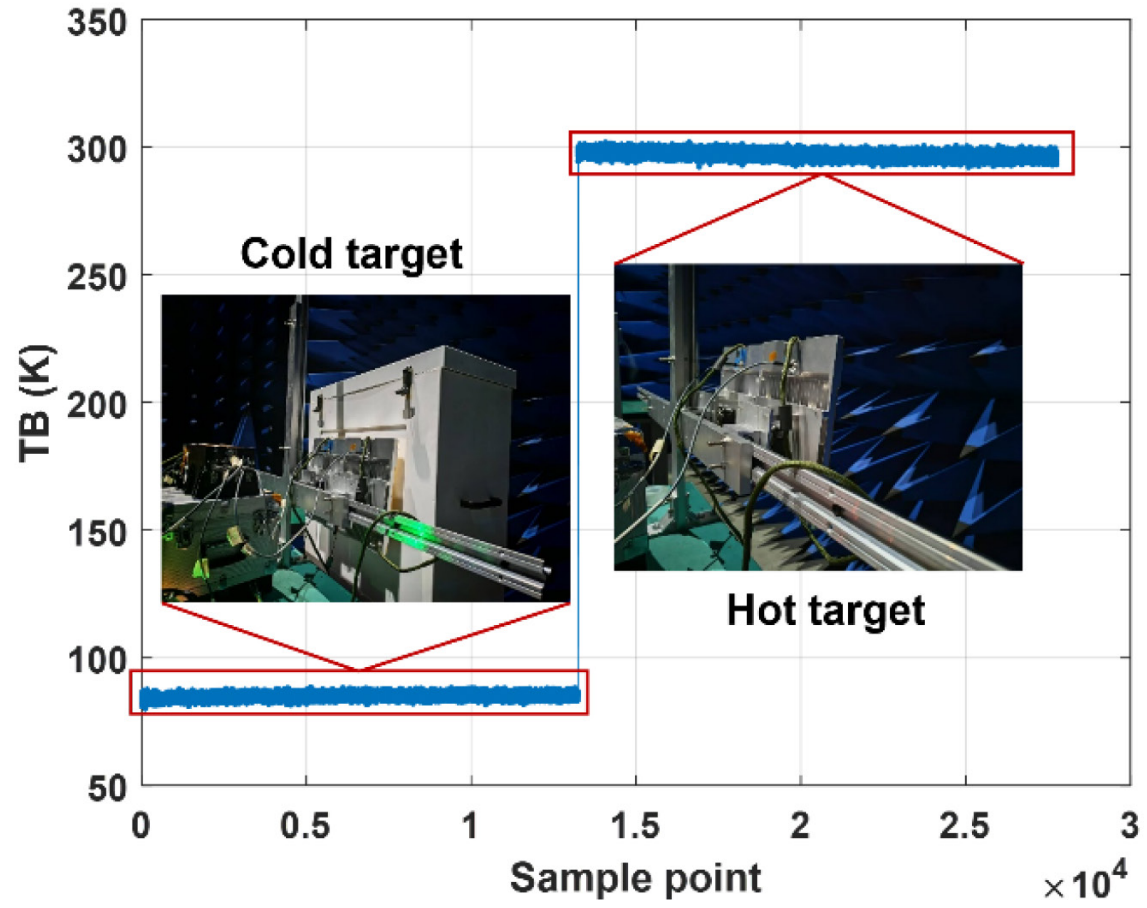

**Fig. 16.** Sensitivity testing results.

The dynamic beam observation experiment was conducted at the outdoor site shown in Fig. 17(a), which is located near a civil air route. The three phased-array antennas formed three interferometric baselines, as illustrated in the inset of Fig. 17(a). During the experiment, the prototype first operated in a wide-area normal scanning mode. Specifically, the beam was electronically steered over a predefined 19×19 grid, and the cross-correlation amplitudes were recorded at each beam position. This operation corresponds to the normal observation mode described in Section II, in which the beam is sequentially steered over multiple pointing directions to form a wide-area response map.

The result of the wide-area scan is shown in Fig. 17(b), where the mean cross-correlation amplitude of the measured baselines is plotted over the scanned beam grid. The high-response region in the lower-left part of the map is mainly attributed to a nearby communication base station rather than to an aircraft target. This result demonstrates that the prototype can electronically scan a two-dimensional beam grid and generate a beam-indexed interferometric response map. Although this experiment does not constitute full passive microwave imaging, it verifies the basic operational procedure required by the normal scanning mode of BF-MIR.

After the wide-area scan, three beam positions were selected for repeated observation. The prototype cyclically switched among Beam Position 1, Beam Position 2, and Beam Position 3, with a dwell time of 300ms at each beam position. This operation corresponds to a selected-beam staring mode, in which a limited number of beam directions are repeatedly observed to monitor temporal variations in specific regions. The aircraft passing through the FOV was used only as an opportunistic moving target to provide a detectable temporal response for this demonstration.

Fig. 17(c) shows the temporal cross-correlation amplitude measured at Beam Position 2. The amplitudes of the three individual baselines, together with their mean value, are plotted as a function of sample index. A transient response is observed around the seventh to ninth samples, which is consistent with the passage of the aircraft through the corresponding beam direction. This response indicates that the prototype can maintain interferometric measurements while dynamically switching among selected beam positions and can capture temporal scene variations within a chosen beam position. Therefore, the experiment provides preliminary validation of the selected-beam staring observation mode of BF-MIR.

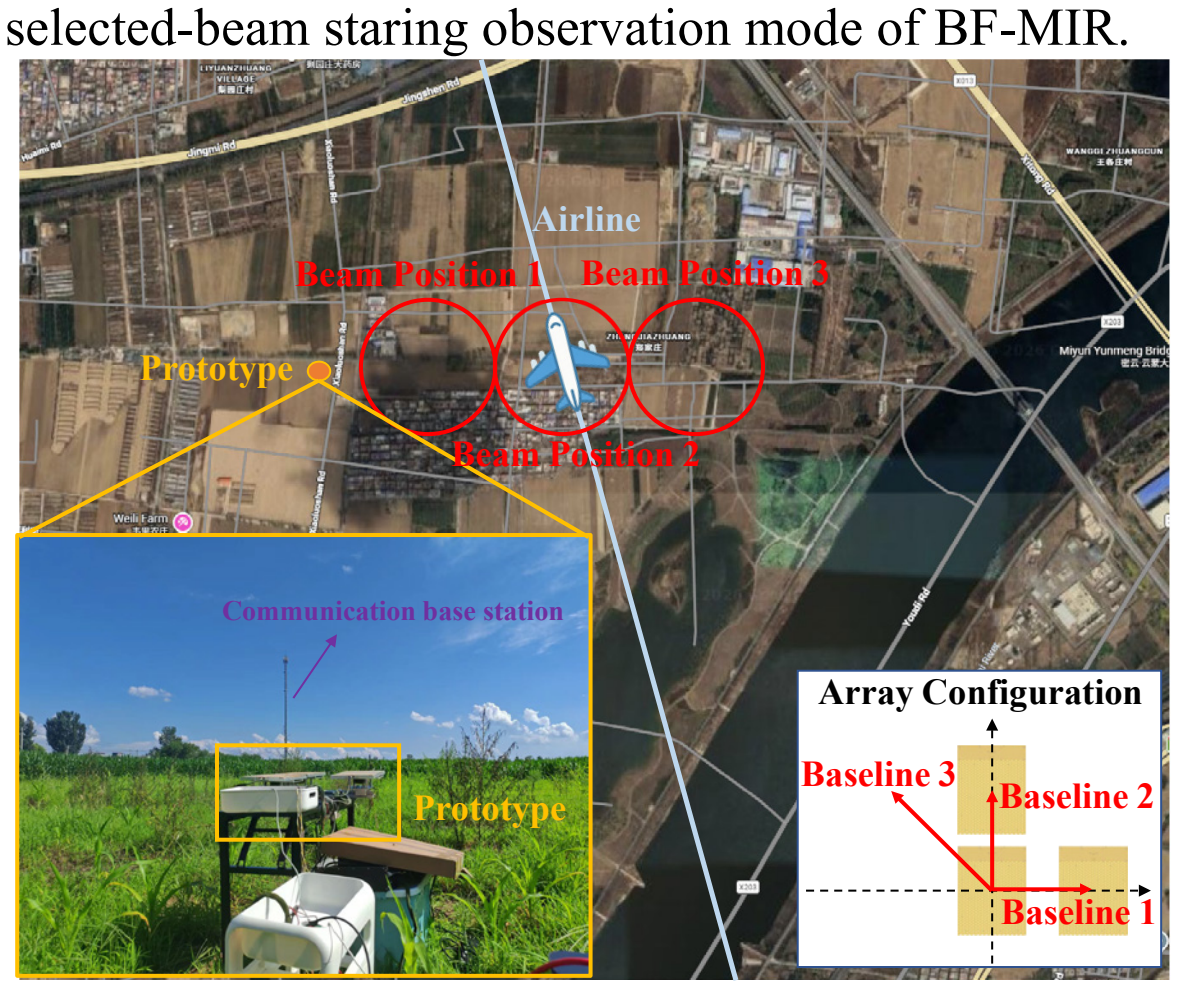

(a)

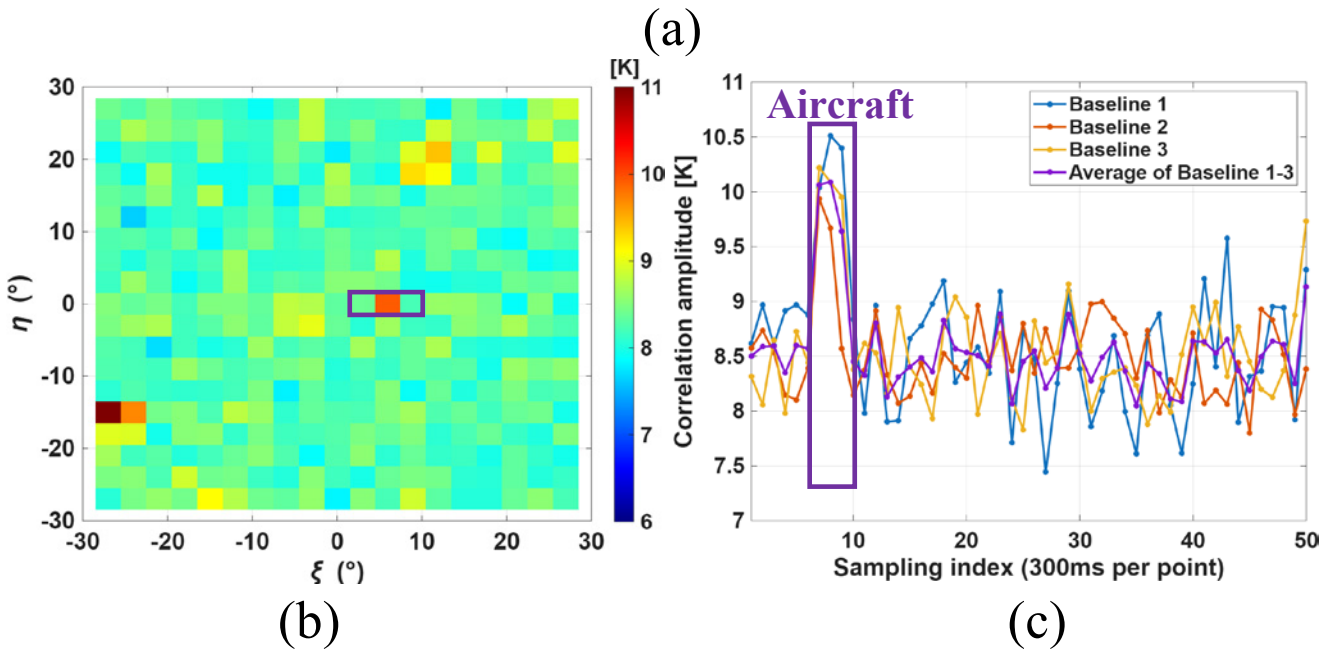

(b) (c)

**Fig. 17.** Detection results for aircraft. (a) Experiment scene; (b) Average correlation amplitude of baseline 1-3; (c) Correlation amplitude of Beam Position 2.

Overall, the results in Fig. 17 demonstrate two basic dynamic beam observation functions of the BF-MIR prototype. The 19×19 beam-grid scan validates wide-area normal scanning, while the repeated three-beam observation validates selected-beam staring under dynamic beam switching. The observed aircraft-induced response is used here only as a convenient scene-change signal, rather than as a quantitative assessment of aircraft-detection performance. Together with the sensitivity check in Fig. 16, these results indicate that the large-aperture beamforming elements used in the prototype can provide stable radiometric measurements during dynamic beam operation.

## V. Discussion

From a system-design perspective, the proposed BF-MIR architecture provides a potential route for extending microwave interferometric radiometry toward higher spatial resolution while maintaining manageable array complexity. By allowing a much larger spatial-frequency sampling interval $d_u$, BF-MIR can synthesize a large aperture with fewer interferometric elements and reduced cross-correlation burden. By employing large-aperture interferometric elements with an increased

ASRF, BF-MIR increases the effective collecting area of each interferometric element, thereby improving radiometric sensitivity, while the narrower element beam acts as a spatial prefilter that suppresses the replicas introduced by spatial-frequency under sampling. In addition, the beamforming capability enables agile beam steering, which compensates for the reduced instantaneous beam-limited coverage associated with narrow element beams.

### A. *System-Level Design Tradeoffs*

The above advantages introduce important system-level tradeoffs. The enlarged $d_u$ makes aliasing an inherent issue in BF-MIR because the spatial-frequency sampling no longer satisfies the conventional Nyquist criterion. The analysis in Section III shows that, under a fixed ASRF, $d_u$ has only a limited influence on aliasing within the main-beam region and can therefore be selected primarily according to element count and correlation-complexity constraints. In contrast, aliasing suppression is mainly governed by antenna-pattern design: increasing the ASRF provides the primary suppression capability, while antenna weighting serves as a secondary optimization tool after sufficient beam narrowing has been achieved. However, ASRF increase reduces instantaneous beam-limited coverage, and stronger weighting reduces aperture efficiency and the usable aperture gain. Therefore, BF-MIR design requires joint consideration of aliasing suppression, beam-limited coverage, aperture efficiency, and application-dependent residual-aliasing tolerance. The E-FOV should accordingly be interpreted as a qualitative, requirement-dependent concept rather than as a fixed threshold-based metric.

### B. *Validation and Scalability of the Proof-of-Concept Prototype*

The preliminary prototype experiments provide initial validation of key operational functions required by BF-MIR. The dynamic beam interferometric phase measurement experiment verifies that consistent fringe-phase measurements can be obtained under beam-steering operation after beam-state-dependent phase-offset compensation. The dynamic beam observation experiment further demonstrates electronically steered wide-area normal scanning and selected-beam staring modes. These results support the feasibility of using beamforming-capable antennas as interferometric elements. Nevertheless, the three-element prototype should be regarded as a preliminary functional demonstrator rather than a full high-resolution imaging system. Further experiments with larger arrays and more representative baseline distributions are required to validate image reconstruction, radiometric performance, aliasing behavior, and dynamic beam operation under practical conditions. The successful demonstration of theses fundamental operations, despite the limited array scale, confirms the basic viability of the BF-MIR concept and encourages further development toward larger implementations.

### C. *Calibration Challenges and Strategies*

Practical implementation of BF-MIR will critically depend on a scalable calibration framework. Compared with conventional fixed-beam MIRs, BF-MIR introduces two major calibration challenges. First, the large physical scale and sparse distribution of the interferometric elements make the calibration of receiver-chain consistency, antenna-pattern errors, and array-geometry uncertainties more difficult. Internal calibration methods developed for conventional MIRs provide an important starting point for correcting receiver-related amplitude and phase mismatches [42], [43], but they are insufficient for beam-dependent antenna-pattern distortions and large-array geometry errors [44], [45]. Second, the beamforming nature of BF-MIR makes the effective gain, phase, and antenna pattern of each interferometric element dependent on the beam-steering state. Therefore, calibration parameters should be modeled as functions of beam pointing rather than treated as fixed channel constants. Similar beam- or geometry-dependent calibration issues have also been recognized in beacon-based calibration, rotating interferometric arrays, and antenna-pattern-error correction studies [45], [46].

A practical calibration strategy for BF-MIR will likely require a hybrid framework combining ground characterization, internal calibration, external reference calibration, beam-state-dependent parameter modeling, and periodic in-orbit updates. Ground calibration should characterize beamforming codebooks, channel transfer functions, antenna patterns, and state-dependent phase offsets over predefined beam states. On-orbit internal calibration can then be used to track short-term receiver drift and maintain receiver-chain consistency, potentially through partial noise injection, dedicated internal references, or cold-sky-assisted amplitude calibration [43], [47]. External calibration remains necessary for residual antenna-pattern, baseline, and beam-state-dependent errors, using references such as well-characterized scenes, beacons, or point-like sources [46], [48]. In addition, processing-level methods, such as model-based reconstruction, de-aliasing, regularized inversion, or data-driven correction, may further reduce residual aliasing or expand the usable imaging region [40], [41]. Developing and experimentally validating these scalable calibration and processing methods remains an important direction for future work.

## VI. Conclusion

This paper proposed a beamforming microwave interferometric radiometer (BF-MIR) architecture for high-resolution passive microwave imaging. By using beamforming-capable antennas as interferometric elements, BF-MIR enables enlarged spatial-frequency sampling intervals and dynamic beam steering, providing a system-level approach to reducing array complexity while maintaining flexible observation capability.

A beamforming interferometric imaging model was established, and the relationships among key system parameters and performance metrics, including spatial resolution, radiometric sensitivity, and beam-limited observation coverage, were analyzed. The results show that enlarged sampling intervals can substantially reduce the required number of interferometric elements and the associated correlation burden for large-aperture passive microwave imaging.

To address the aliasing introduced by large sampling intervals, a Shift-Accumulate method was developed for image-domain aliasing analysis. The analysis showed that, under a fixed ASRF, the sampling interval has only a limited effect on aliasing within the main-beam region, whereas antenna-pattern design plays the dominant role. Increasing the ASRF provides the primary aliasing suppression capability, and antenna weighting can be used as a secondary optimization after sufficient beam narrowing has been achieved.

A three-element proof-of-concept prototype was built to provide preliminary experimental validation of the proposed architecture. The dynamic beam interferometric phase measurement experiment verified consistent fringe-phase measurement under beam-steering operation after beam-state-dependent phase-offset compensation. The dynamic beam observation experiment further demonstrated wide-area normal scanning and selected-beam staring modes.

Future work will focus on large-scale array implementation, end-to-end imaging validation, and dedicated calibration and processing methods for dynamic beam interferometric radiometry.